\documentclass[12pt,preprint]{aastex}

\slugcomment{to appear in the {\it Astronomical Journal}}

\shorttitle{Parallaxes from CTIOPI --- MOTION Stars}
\shortauthors{Jao et al.}

\begin{document}

\title{The Solar Neighborhood XIII: Parallax Results from the CTIOPI
0.9-m Program -- Stars with $\mu$ $\ge$ 1\farcs0/year (MOTION Sample)}

\author{Wei-Chun Jao\altaffilmark{1}, Todd J. Henry\altaffilmark{1},
John P. Subasavage\altaffilmark{1} and Misty A. Brown\altaffilmark{1}}

\affil{Georgia State University, Atlanta, GA 30302-4106}
\email{jao@chara.gsu.edu, thenry@chara.gsu.edu,
subasavage@chara.gsu.edu, brown@chara.gsu.edu}

\author{Philip A. Ianna\altaffilmark{1}, Jennifer L.
Bartlett\altaffilmark{1}} \affil{University of Virginia,
Charlottesville, VA 22903} \email{pai@virginia.edu,
jlb2j@virginia.edu}

\and 

\author{Edgardo Costa\altaffilmark{1}, Ren\'{e}
A. M\'{e}ndez\altaffilmark{1}} \affil{Universidad de Chile, Santiago,
Chile} \email{costa@das.uchile.cl, rmendez@das.uchile.cl}

\altaffiltext{1}{Visiting Astronomer, Cerro Tololo Inter-American
Observatory.  CTIO is operated by AURA, Inc.\ under contract to the
National Science Foundation.}

\begin{abstract}

We present the first set of definitive trigonometric parallaxes and
proper motions from the Cerro Tololo Inter-American Observatory
Parallax Investigation (CTIOPI).  Full astrometric reductions for the
program are discussed, including methods of reference stars selection,
differential color refraction corrections, and conversion of relative
to absolute parallax.  Using data acquired at the 0.9-m at CTIO, full
astrometric solutions and $VRIJHK_s$ photometry are presented for 36
red and white dwarf stellar systems with proper motions faster than
1\farcs0/yr.  Of these, thirty three systems have the first ever
trigonometric parallaxes, which comprise 41\% of MOTION systems (those
with proper motions greater than 1\farcs0/yr) south of $\delta$ $=$ 0
that have no parallaxes.  Four of the systems are new members of the
RECONS 10 pc sample for which the first accurate trigonometric
parallaxes are published here: DENIS J1048-3956 (4.04 $\pm$ 0.03 pc),
GJ 1128 (LHS 271, 6.53 $\pm$ 0.10 pc), GJ 1068 (LHS 22, 6.97 $\pm$
0.09 pc), and GJ 1123 (LHS 263, 9.02 $\pm$ 0.16 pc).  In addition, two
red subdwarf-white dwarf pairs, LHS 193AB and LHS 300AB, are
identified.  The white dwarf secondaries fall in a previously
uncharted region of the HR diagram.

\end{abstract}

\keywords{astrometry --- stars: distance --- photometry --- solar
neighborhood --- stars: high proper motion stars --- white dwarfs}

\section{Introduction}

The first ever stellar trigonometric parallax was reported by
F. Bessel in 1838 for 61 Cygni, after a ``race'' in which he narrowly
defeated F. G. Wilhelm Struve and T. Henderson, who published the
parallaxes for Vega and $\alpha$ Centauri, respectively, in the next
year.  Since then, trigonometric parallax measurements have provided
one of the most important parameters for understanding stellar
astronomy --- distance --- and provide one of the sturdiest rungs on
the cosmic distance ladder.  Trigonometric parallaxes are used to
derive the intrinsic luminosities of stars, calculate accurate masses
for binary system components, and answer questions about stellar
populations and Galactic structure.  In addition, the solar
neighborhood is mapped out via trigonometric parallaxes, and these
nearby objects provide the brightest examples of many stellar types,
supplying the benchmarks to which more distant stars are compared.

Two of the most important parallax references are the Yale Parallax
Catalog \citep{YPC} and Hipparcos Catalog \citep{Hipparcos}.
Combined, they offer $\sim$120,000 parallaxes both from space and
ground observations.  Of these, 92\% of trigonometric parallaxes are
from the Hipparcos mission.  However, because of the relatively bright
magnitude limit of Hipparcos, many nearby stars candidates were
excluded.  Consequently, the faint members of the solar neighborhood
are under-presented, and these faint red, brown, and white dwarfs are
the objects targeted by recent trigonometric parallax efforts,
including the one discussed in this paper.  Recent results for nearby
red and brown dwarfs include the efforts of \citet{Ianna1996},
\citet{Tinney1995, Tinney2003}, \citet{Dahn2002}, and
\citet{Vrba2004}, which together have provided $\sim$130 total
ground-based parallaxes since 1995.

\section{The CTIOPI Effort and Sample}

In order to recover ``missing'' members in the solar neighborhood, the
Research Consortium On Nearby Stars (RECONS) group is currently
carrying out a southern sky parallax survey known as Cerro Tololo
Inter-American Observatory Parallax Investigation (CTIOPI).  The
primary goals of CTIOPI are to discover and characterize nearby red,
brown, and white dwarfs that remain unidentified in the solar
neighborhood.  This program was selected as a NOAO Survey Program, and
observations commenced in 1999 August.  CTIOPI used both 0.9-m and
1.5-m telescopes during the NOAO Survey Program, and has continued on
the 0.9-m as part of the SMARTS (Small and Moderate Aperture Research
Telescope System) Consortium beginning in 2003 February.  The RECONS
team at Georgia State University is responsible for data reduction for
the 0.9-m program, while data from the 1.5-m program is being analyzed
by E. Costa and R. M\'{e}ndez of the Universidad de Chile in Santiago.
The extended 0.9-m program has recently surpassed 400 systems on the
observing list, while the final 1.5-m program included $\sim$50
systems that were fainter (and for which less observing time was
awarded).

Most of the target stars (hereafter, the ``pi stars'') are selected
for CTIOPI because available astrometric (e.g, high proper motion),
photometric, or spectroscopic data indicate that they might be closer
than 25 pc.  In the 0.9-m program, roughly 95\% of the pi stars are
red dwarfs and the remainder are white dwarfs.  The fainter brown
dwarf candidates were included in the 1.5-m program.  In all,
$\sim$30\% of the 0.9-m targets are members of what we call the MOTION
sample --- stellar systems having proper motions of at least
1\farcs0/yr.  This paper describes the first definitive astrometric
results of CTIOPI, focusing on the results for 36 MOTION systems.

\section{Observations}
\subsection{Astrometry Observations}

The 0.9-m telescope is equipped with a 2048 $\times$ 2048 Tectronix
CCD camera with 0\farcs401/pixel plate scale \citep{Jao03}.  All
observations were made using the central quarter of the chip, yielding
a 6\farcm8 square field of view, through $V_{J}$, $R_{KC}$ and
$I_{KC}$ filters (hereafter, the subscripts are not
given)\footnote{Subscript: J $=$ Johnson, KC $=$ Kron-Cousins. The
central wavelengths for $V_{J}$, $R_{KC}$ and $I_{KC}$ are 5475\AA,
6425\AA~and 8075\AA, respectively.}.  The dewar containing the CCD
camera is mounted on the telescope with columns oriented in the
north-south direction.  A slight rotation relative to the sky is
possible because of instrument flexure and repositioning during
telescope maintenance.  This rotation angle can be calibrated, as
discussed in section~\ref{astro.initial.steps}.

The observing procedures employed during CTIOPI mimic those used in
University of Virginia southern parallax program at the Siding Spring
Observatory in Australia, led by P. Ianna, who is a member of the
CTIOPI team.  When a star is observed for the first time, exploratory
exposures are taken in the $VRI$ filters to find a suitable set of
reference stars in the field.  The parallax filter and position of the
field are selected to balance the brightness of the pi star with
available potential reference stars, and that filter is used for all
subsequent parallax frames.  Because most of our pi stars are nearby
star candidates, they are brighter than most of the field stars.  We
attempt to place the pi star on the chip so that 5 to 15 field stars
of adequate flux can be included.  Typically, a good reference star is
not more than twice as bright as the pi star (in the few cases when
the pi star is not the brightest star in the field), but has at least
1000 peak counts during a typical parallax exposure.  

Bias frames and dome flats are taken at the beginning of each night to
allow for basic data reduction calibration.  Parallax observations are
usually made within $\pm$30 minutes of a pi star's transit in order to
minimize the corrections required for differential color refraction
(DCR) corrections, which are discussed in section~\ref{dcr}.  A few
faint pi stars are observed with a wider hour angle tolerance because
frame acquisition takes longer.  Exposure times for parallax frames
typically provide a peak of $\sim$50,000 counts for the pi star
(saturation occurs at 65,535 counts), in an effort to maximize the
number of counts available for pi star and reference star centroiding.
Usually, 3--10 frames are taken in each visit, depending primarily on
the exposure time required.  Multiple frames are taken to reduce the
errors on the pi star and reference star positions at each observation
epoch.  The typical set of observations required to determine a final
parallax and proper motion includes four seasons of observations
carried out over at least 2.5 years (further details in
section~\ref{pi.quality.check}).

\subsection{$VRI$ Photometry Observations}
\label{sec:phot.reduce}

The $VRI$ photometry reported here was acquired at the CTIO 0.9-m
utilizing the same instrument setup used for the astrometry frames.
All of the results are from observations taken between November 1999
and September 2004.  As with the astrometry observations, bias and
dome flat images were acquired nightly and used for basic science
frame calibration.

Most pi stars were observed at $\sec z$ $<$ 1.8 or less (a few were
between 1.8 and 2.0 airmasses because of extreme northern or southern
declinations).  Various exposure times were used to reach S/N $>$ 100
for pi stars in each of the $VRI$ filters.  Standard star fields with
averagely total 10 stars from \citet{Landolt1992} and/or E-regions
from \citet{Graham1982} were observed several times each night to
derive transformation equations and extinction curves.  In addition,
one or two very red standard stars with $V-I >$ 3.7 were observed in
order to derive extinction coefficients for stars with a wide range of
colors.  Typically, a total of 4--5 standard star fields were observed
2--3 times each per night.

\section{Astrometry Reductions}
\subsection{Initial Data Processing, Reference Star Selection, and Trail Plate Selection}
\label{astro.initial.steps}

The basic data reduction for the astrometry CCD frames includes
overscan correction, bias subtraction and flat-fielding, performed
using a customized IRAF package called {\em redpi} (because our pi
stars are primarily red dwarfs).  After processing the raw data,
frames are sorted into storage directories by object until there are
enough parallax frames and time coverage to derive a reliable
astrometric solution, typically at least 40 frames over at least two
years.  When sufficient frames are available, {\em SExtractor}
\citep{sextractor} is used to determine centroids for each pi star and
a set of reference stars that is chosen using the following general
guidelines:

\begin{enumerate}

\item A single frame is selected that was taken using the parallax
filter.  The seeing is required to be better than 1\farcs5 and images
must have ellipticity less than $\sim$20\%.

\item Five to 15 reference stars in the field are selected that evenly
surround the pi star in order to determine reliable plate
rotation, translation, and scaling coefficients.

\item Each reference star must have a complete set of $VRI$
photometry, which is required for DCR corrections and the conversion
of relative to absolute parallax.

\item Using the IRAF {\it imexam} task, each reference star is checked
to make sure that it is not a resolved binary or galaxy.

\item Ideally, all of the reference stars are have peak counts above
1000, although some fields require the selection of fainter stars in
order to have a sufficient number of reference stars.

\end{enumerate}

\noindent After the first round of parallax reductions, each reference
star is reexamined for a sizable parallax or proper motion, and
removed from the reference field if necessary.

In order to calculate the parallax factors, accurate coordinates and a
value for the Earth to Solar System barycenter distance need to be
known.  The coordinates used for parallax factor calculations are
extracted from the Two Micron All Sky Survey (2MASS) All-Sky Point
Source Catalog via OASIS.  Because these objects are high proper
motion stars, all of them were manually identified by comparison with
finding charts instead of retrieving data blindly by setting a
searching radius around a given RA and DEC.  The coordinates listed in
Table~\ref{tbl:pi.result} for the pi stars have been shifted to epoch
2000 using available proper motion measurements, primarily
\citet{LHS}, instead of the epoch at which the images were taken by
2MASS.  To compute an accurate distance from the Earth to the Solar
System barycenter at the time of observation, the JPL ephemeris DE405
is used.

Before calculating the parallax and proper motion of the pi star using
frames taken at many epochs, a single ``trail plate'' is selected as a
fundamental reference frame to which all other images are compared.
This trail plate is used to remove any rotation, translation, and
scaling changes between frames.  A customized program organizes the
set of frames used during the reductions for a particular field, is
run to calculate the hour angle, parallax factors, and FWHM of images
for each frame, and a trail plate is selected using the results, using
the following criteria:

\begin{enumerate}

\item All reference stars and pi star(s) have peak counts less than
65500 and greater than 100.

\item All reference stars and pi star(s) have ellipticity less than
$\sim$20\%.

\item All the reference stars and pi star(s) have FWHM less than
2\farcs5.  This criterion has been relaxed relative to the frame used
for the initial selection of reference stars (when 1\farcs5 is the
limit) in order for the trail plate to be nearly coincident with the
meridian.

\item The hour angle is within 2 minutes of the meridian at the
midpoint of the integration.  A frame taken very near the meridian
provides a trail plate with minimal DCR.

\end{enumerate}

\noindent Usually, the definitive trail plate is the one having the
smallest hour angle and best seeing of the frames available.

The rotation angle of the trail plate is calculated relative to the
Guide Star Catalog 2.2 (GSC2.2) using WCSTools/imwcs\footnote{WCSTools
is available at
\url{http://tdc-www.harvard.edu/software/wcstools/}.}. Our parallax
images are usually deeper than GSC2.2, so stellar objects with
apparent magnitudes brighter than 18.0 and FWHM smaller than 2\farcs5
(but larger than 0\farcs6 to avoid centroiding on cosmic rays and bad
pixels) are used.  Once the rotation angle is determined, the parallax
factors and centroids for all reference stars and pi stars on the
trail plate are recalculated and used as the fundamental reference
frame.

\subsection{Differential Color Refraction Corrections}
\label{dcr}

DCR corrections are required because the pi star and reference stars
are not of identical color; therefore, their positions as seen from
underneath Earth's atmosphere shift relative to one another because of
different, but calibrateable amounts of refraction.  Although most of
our parallax observations suffer minimal DCR because they are made
within 30 minutes of the meridian, sometimes frames are taken far
enough from the meridian that it is advantageous to make DCR
corrections, e.g. for important targets observed in non-ideal
observing seasons and in cases when the total number of available
frames can be boosted by utilizing photometry frames taken in the
parallax filter.  Different observing and reduction methods used to
measure DCR have been discussed by \citet{Monet1992},
\citet{Tinney1993}, \citet{Stone1996}, and \citet{Stone2002}.  Here we
use both the theoretical methods proposed by \citet{Stone1996} and the
empirical methodology proposed by \citet{Monet1992} to measure DCR for
the CTIOPI program, and to make final corrections during astrometric
reductions.

DCR calibration observations for CTIOPI were made during four
photometric nights in December 2002 using the 0.9-m telescope at CTIO.
This is the identical combination of telescope, filters, and CCD
camera used during the parallax program.  Ten different fields spread
from zenith to low altitude that contained blue ($V-I = 0.57$) to red
($V-I = 3.22$) stars were selected and observed through the $V$, $R$
and $I$ filters.  Ten fields were each observed up to five times per
night through hour angles of $\pm$4 hours.  Exposure times were chosen
to be the same as used in the parallax program for each field so that
the faintest reference stars could be analyzed for DCR.  In total, 72
stars were included in the final DCR analysis.  Although refraction
is, in general, a function of temperature, pressure, and humidity, due
to the stable observing conditions throughout this run, these factors
can be ignored, as discussed in \citet{Monet1992} and
\citet{Stone2002}.

In order to provide a zero point reference frame for the DCR
calculation, one set of images must be taken when the field transits.
In other words, there is no refraction in the RA direction, and we
assign zero refraction in the DEC direction, when the hour angle is
zero.  The components of refraction in the RA and DEC directions,
$R_{m}Z_{x}$ and $R_{m}Z_{y}$ respectively (where $R_{m}$ is the mean
refraction), are given by

\begin{eqnarray}
(\alpha-\alpha^{\prime})\cos\delta  =\frac{R_{m}\sin HA
\cos\phi}{\cos\zeta}= R_{m}\cos\phi\sin HA\sec\zeta=R_{m}Z_{x}  \\
\delta-\delta^{\prime} = R_{m} S \sin\phi\sec\delta(\sec\zeta-\sec(\phi-\delta))=R_{m}Z_{y},
\label{eqn:dcr.correction}
\end{eqnarray}

\noindent where ($\alpha$, $\delta$) are the coordinates without
atmospheric refraction and ($\alpha^{\prime}$, $\delta^{\prime}$) are
the coordinates after atmospheric refraction.  The angle $\phi$ is the
latitude of the observing site, HA is the hour angle of a given star
and $\zeta$ is its zenith distance.  As discussed in
\citet{Monet1992}, $S$ merely represents the sign of the declination
term, i.e.~$S = 1$ if $(\phi-\delta)\geq 0$ and $S = -1$ if
$(\phi-\delta)< 0$.  These empirical measurements assume that $R_{m}$
is a polynomial function of $V-I$ color (see also \citet{Monet1992}).
We have determined the $V$, $R$, and $I$ magnitudes for all 72 stars
in the ten fields used to calculate the DCR so that each filter can be
calibrated against the $V-I$ color (thereby producing three sets of
equations as shown in the next section).

\subsection{The Final DCR Model for CTIOPI}

The images taken for the DCR model were reduced in a manner identical
to the parallax frames, as discussed in
section~\ref{astro.initial.steps}.  The three $VRI$ frames having the
smallest hour angle were selected as trail plates assumed to have no
DCR.  Plate constants are calculated using the GaussFit\footnote{This
is a program for least squares and robust estimation that is available
from the Hubble Space Telescope (HST) Astrometry Team {\em
ftp://clyde.as.utexas.edu/pub/gaussfit/manual/}.}  program
\citep{Jefferys1987}.  Six plate constants are derived so that field
rotation, translation, and scaling can be removed (see
section~\ref{astro.many.epochs}).  We ignore any effects of source
proper motion or parallax during the four nights of DCR observations
because they are negligible on that time scale.  Consequently, after
calculating the plate constants, the only shifts in stellar centroids
are because of atmospheric refraction.  The amount of centroid shift
from the trail plate in the X direction is a direct measure of the
refraction, as represented by the quantity on the far left side of
Equation 1.  \citet{Monet1992} (see their Figure 2) showed that because
the refraction in the Y direction has been defined as shown in
Equation 2 (effectively removing any shift in the Y direction for zero
hour angle), the X shift ($R_{m}Z_{x}$) will have more variation than
the Y shift ($R_{m}Z_{y}$) when the hour angle is different from
zero. Therefore, we concentrate on the RA direction to determine the
empirical polynomial function for $R_{m}$.

To determine the functional form of $R_{m}$, first the hour angle and
$Z_{x}$ for every useful star in the ten DCR calibration fields are
derived.  Then, based on the centroid shift and $Z_{x}$ for various
stars observed during the observing run, the slope of $R_{m}$ versus
$Z_{x}$ can be found.  Figure~\ref{fig:dcr.slope} shows an example for
LHS 158 and a reference star in the field.  The field was observed
from 1.47 hours east of the meridian to 3.71 hours west, including
eight sets of $VRI$ observations at different hour angles.  A linear
fit, whose slope is $R_{m}$, was made for each filter to each of the
72 stars selected in the ten fields in order to provide an ensemble of
values, $R_{m}$, as a function of $V-I$ color.

We set the zero point for DCR to be $R_{m} = 0$ when $V-I = 0$,
thereby defining a star of that color to show no DCR, while all other
stars' DCR is measured relative to that.  The $0^{th}$ order
coefficient for each field is slightly different from the others
because there is rarely a star with $V-I = 0$ in a frame, but the
offset can be computed by a least squares fit for a polynomial
function to all stars that are present in a given field.  By combining
the $R_{m}$ slopes and the $V-I$ values for all 72 stars, we generate
the plots in Figure~\ref{fig:dcr.fit.plot}, showing the empirical fits
with solid curves.

The mean empirical DCR functions\footnote{Different orders of
polynomial fits were calculated for each filters. The ones with
reasonable slope and points distribution are given.} for three
different filters are given by:

\begin{eqnarray}
R_{m,V} =-0.0407(V-I)+0.00941(V-I)^{2}, \nonumber \\
R_{m,R} =-0.0417(V-I)+0.0482(V-I)^{2}-0.0245(V-I)^{3}+0.0036(V-I)^{4}, \nonumber \\ 
R_{m,I} =+0.0007(V-I).
\label{eqn:dcr.function}
\end{eqnarray}

The theoretical curves for all three filters were also calculated
using the model from \citet{Stone1996} and are shown in
Figure~\ref{fig:dcr.fit.plot} as dashed lines\footnote{The FORTRAN
code used to generate the curves was kindly provided by M. Begam from
the Siding Spring Observatory parallax project, led by P. Ianna.}.  A
hypothetical field at DEC $=$ $-$26, ``observed'' during a night with
temperature $T = 12^{\circ}C$ and $40\%$ humidity was chosen to
generate the model curves.  These conditions are similar to those
encountered during CTIOPI observations.  Twelve stars with spectral
types of A0 V to M5 V were selected for the model, and were
``observed'' at positions that were 0 to 3 hours from the meridian.

As expected, Figure~\ref{fig:dcr.fit.plot} shows that $I$ band has the
least DCR of the three filters.  In all three filters the average
difference between the model and the empirical curve is always less
than 6 mas for stars with $V-I <$ 3.2.  Because our DCR sample is
deficient in very red stars, the difference between the empirical and
theoretical curves increases at the red end of the $R$ band
calibration.  Note that when the stellar color is redder than $V-I =
2.6$, stars observed through the $R$ filter will actually experience
more DCR than they will when observed through the $V$ filter.  This
result can be explained because we are discussing ``differential''
color refraction among a set of stars.  At a given position in the
sky, the amount of refraction is caused primarily by two factors ---
how photons in a given filter bandpass are refracted by the Earth's
atmosphere, and how the number of photons changes within the bandpass,
i.e. the slopes of the various stellar spectra.  In the case of the
$VRI$ filters, the $R$ filter has the largest amount of refraction for
the reddest stars because both factors are important, whereas in the
$V$ band the slopes of the stellar spectra do not change much for very
red stars, and in the $I$ band, the atmosphere does not refract the
photons significantly, regardless of a star's color. Consequently, the
DCR correction for each star can be made by obtaining its $V$ and $I$
photometry and applying Equation 1 to 3.

A valuable comparison of astrometry reductions is shown in
Figure~\ref{fg:dcr.gj1061}, in which results from two reductions are
presented for the same data --- one with and one without DCR
corrections.  A series of high hour angle measurements were taken in
mid-2003 to test our DCR protocol. The effects of DCR corrections are
clearly seen when comparing these two panels. In the case of no DCR
corrections (the two plots on the left), the X direction residuals
show a very deep ``valley'' and the Y direction residuals show a large
scatter.  After the DCR corrections are applied, the X residuals
flatten out, and the Y residuals are reduced and more symmetric around
zero. The standard deviations for X and Y residuals drop from 12.5 and
8.6 mas to 5.6 and 8.4 mas, respectively, when DCR corrections are
made.  The larger reduction in the X direction is expected because the
differential refraction is more significant in the RA direction.

\subsection{Least Squares Reduction of Images Taken at Many Epochs}
\label{astro.many.epochs}

Once DCR corrections are incorporated into the data reduction
pipeline, the positions of a pi star and a set of reference stars can
be accurately computed for an ensemble of frames, with each frame in
the ensemble being compared to the trail plate.  The relationship
between a frame and the trail plate is based on the measured positions
of reference stars only (not the pi star).  A new set of coordinates
for each reference star is derived as a function of the trail plate
coordinates and a set of constants:

\begin{eqnarray}
\xi=Ax+By+C, \nonumber \\
\eta=Dx+Ey+F,
\label{eq:plate.constant}
\end{eqnarray}

\noindent where $(x,y)$ are the original coordinates of a reference
star, A--F are the {\em plate constants}, and $(\xi,\eta)$ are the
coordinates after the transformation.  This six-constant model allows
for a different scale in both the X and Y directions, compensates for
different amounts of translation in both directions, and includes a
correction for any instrument rotation.  The higher order plate
constants --- radial distortion: $Rx(x^{2}+y^{2})$, coma: $Smx$ (m is
magnitude), and decentering distortion: $P(x^{2}+y^{2})$
\citep{Eichhorn1974} --- are not included in the current calculations
because parallax results from our standard stars are within 2$\sigma$
of all other observations and no systematic differences are seen
(discussed in section~\ref{pi.quality.check}).

Analysis of the stellar path of the pi star must take into account
both proper motion and parallax, but each reference star also
experiences both motions on the sky.  Because accurate proper motions
and parallaxes are rarely known for reference stars, we assume that
the reference grid has $\sum_{i}\pi_{i} = 0$ and $\sum_{i}\mu_{i} = 0$
(\citealt{Altena1986}, \citealt{Benedict1999}).  Hence, the set of
constants for each frame outlined in Equation 4 above is expanded to
include the reference star motions, resulting in an expanded set of
equations:

\begin{eqnarray}
\xi_{1}^{t} = A^{1}x_{1}^{1}+B^{1}y_{1}^{1}+C^{1}+\mu_{x1}T+\pi_{1} P_{\alpha 1}, \nonumber \\
\xi_{2}^{t} = A^{1}x_{2}^{1}+B^{1}y_{2}^{1}+C^{1}+\mu_{x2}T+\pi_{2} P_{\alpha 2}, \nonumber \\
\xi_{3}^{t} = A^{1}x_{3}^{1}+B^{1}y_{3}^{1}+C^{1}+\mu_{x3}T+\pi_{3} P_{\alpha 3},           \\
... \nonumber\\					    	       	
\xi_{n}^{t} = A^{1}x_{n}^{1}+B^{1}y_{n}^{1}+C^{1}+\mu_{xn}T+\pi_{4} P_{\alpha n}, \nonumber
\end{eqnarray}

\noindent where superscripts indicate frame numbers, subscripts
indicate the identification numbers of reference stars, the product
$\mu$$T$ is the star's total proper motion relative to the date of the
trail plate, the product $\pi$$P$ is the parallax offset from the date
of the trail plate ($P_{\alpha}$ is the parallax factor in RA), and
$\xi_{n}^{t}$ represents the x coordinate for the trail plate.  The
plate constants, $A$, $B$, and $C$ can be calculated from these
equations using least squares methods which are constrained by the
conditions of reference star parallaxes and proper motions summing to
zero.  A similar set of equations is obtained for the y coordinate
(plate constants $D$, $E$, and $F$ in Equation 4).  After the plate
constants and reference star values for $\mu$ and $\pi$ are acquired,
$\mu$ and $\pi$ (and their errors) are computed for the pi star.

The least squares calculation is run using Gaussfit (discussed in
section 4.3), which typically requires three iterations to minimize
$\chi$$^2$.  The image quality of each frame and the reliability of
reference stars are determined using the results of the initial run of
Gaussfit.  At this stage, reference stars with high proper motion,
large parallax, large centroid residuals, or high photometric parallax
are deleted.  Entire frames with high residuals are also removed.  The
Gaussfit program is then run again to derive the final pi star $\mu$
and $\pi$ values.

\subsection{Conversion from Relative Parallax to Absolute Parallax}

What we have measured reflects the parallax of the pi star relative to
the set of reference stars is the relative trigonometric parallax,
$\pi$. As discussed in \cite{Altena1974} and \cite{Altena1988}, there
are (at least) three different ways to convert this {\em relative
parallax} to the~{\em absolute parallax}, which is a measure of the
true distance to the pi star --- using statistical methods,
spectroscopic parallaxes, or photometric parallaxes for the reference
stars.

Statistical methods rely on a model of the Galaxy for the disk and
halo.  By adopting a Galactic model and knowing the apparent
magnitudes and Galactic coordinates of the reference stars, parallaxes
can be estimated for the reference stars.  No reference star color
information is used.  For example, \cite{Altena1988} concludes that
faint halo stars have ($14.5 < V < 15.5$) with a narrow distribution
in their parallaxes for fields near the north Galactic pole.  However,
bright disk stars ($10.5 < V < 11.5$) exhibit a wide range of
parallaxes.  Therefore, faint reference stars have smaller mean
parallaxes and require a small correction for the relative to absolute
parallax conversion, while brighter reference stars require larger
corrections.  As discussed in section~\ref{astro.initial.steps}, the
reference stars chosen for CTIOPI are the brightest available in the
pi star fields (in order to obtain better centroids), so we do not use
a statistical methodology for the conversion of relative to absolute
parallax.

Using spectroscopic parallaxes is arguably the most reliable method to
determine the correction from relative to absolute parallax because
the spectral type and luminosity class of every reference star are
determined.  This allows us to distinguish main sequence stars from
giants and subdwarfs, and to apply correct $M_{V}-color$ relations for
each class of star.  However, this method requires a significant
amount of observing time, and is not practical for CTIOPI, in which
several hundred stars with $\sim$10 reference stars each are observed.

Instead, we use the photometric parallax method to convert the pi
star's relative parallax to its absolute parallax.  $VRI$ magnitudes
for the pi star and all reference stars have already been acquired for
the DCR corrections, so the same data can be used to estimate a
parallax of each reference star.  However, because of the lack of
information about the luminosity class of these stars, these
corrections assume that all of the reference stars are main-sequence
stars.  Additional corrections for the contamination by giants or
galactic reddening have not been included because such corrections are
anticipated to be much smaller than the typical errors on the final
parallaxes.

The fundamental relations between $M_{V}$ and color used in CTIOPI are
based on the sample of stars within 10 pc (Henry et al.~1997, 2004).
Close multiple stars, subdwarfs, evolved stars, and stars with poor
trigonometric parallaxes have been deleted from this sample to provide
reliable $M_{V}-color$ relations.  Three different colors, $V-R$,
$V-I$, and $R-I$, are used to calculate the mean photometric parallax
for each reference star.  The error on the photometric parallax for an
individual star is taken to be the average difference between the mean
photometric parallax and the parallax from each color.  The weighted
mean photometric parallax of the entire set of reference stars is then
calculated, and represents the final correction from relative to
absolute parallax.  The error in the final correction is determined
from

\begin{equation}
\frac{err_{1}/\pi^{phot}_{1}+err_{2}/\pi^{phot}_{2}+...+
err_{n}/\pi^{phot}_{n}}{n}\times \pi_{weighted-mean},
\end{equation} 

\noindent where $n$ is the number of reference stars, $err$ is the
photometric parallax error of each star and $\pi_{weighted-mean}$ is
the weighted mean photometric parallax of the ensemble of reference
stars.  We note that the mean absolute parallax correction for all 36
MOTION stars in Table~\ref{tbl:pi.result} is 1.47 $\pm$ 0.17 mas.

\section{Photometry Reductions}

The same {\em redpi} package discussed in
section~\ref{astro.initial.steps} is used to process the raw
photometry data.  Stars of interest, including pi stars, reference
stars, and photometric standard stars, are tagged and enclosed in an
aperture with a 7\arcsec~radius if there are no nearby background
stars that might contaminate the photometry.  A 7\arcsec~radius
aperture was used for the standard stars in order to match the
aperture typically used by \citet{Landolt1992}.  After removing cosmic
rays, the instrumental magnitude is determined by summing all of the
counts for pixels falling in the aperture.  In the few cases where a
contaminating source is within the 7\arcsec~aperture, an aperture
correction is performed.  A sky annulus with 20\arcsec~inner radius
and 3\arcsec~width was applied to calculate the sky background counts.

The transformation equation for apparent magnitude is

\begin{equation}
m_{standard}=m_{inst}+a_{1}+a_{2}(AM)+a_{3}(color)+a_{4}(color)(AM),
\end{equation}

\noindent where $m_{inst}$ is the instrumental magnitude from {\em
IRAF/DAOPHOT}, $a_{1}$ through $a_{4}$ are the transformation
coefficients, $color$ is the color term (which may have various
permutations using $VRI$ magnitudes), $AM$ is the airmass and
$m_{standard}$ is the standard magnitude from \citet{Landolt1992}.
The {\em IRAF/fitparam} task is used to compute these coefficients via
a least squares method.  To generate the final $VRI$ magnitudes on the
Johnson-Kron-Cousins system, the transformation equation is applied
using a custom-made Perl task. The advantage of this Perl script over
the {\em IRAF/evalfit} task is that the output file contains not only
the $VRI$ apparent magnitudes, but image names, magnitude errors, and
the date of data reduction.  These output files are then concatenated
into a large master photometry database for future access.

\section{Parallax Results}
\subsection{Parallax Results for Calibration Stars}
\label{pi.quality.check}

Seven parallax standard stars were selected to check the reliability
of CTIOPI results.  They were selected so that different parts of the
sky were represented.  All but one, LHS 1777, are within 10 pc and
have final parallax determinations with more than 60 frames spanning
more than 2.5 years.

The trigonometric parallax results for these stars from CTIOPI and
other sources are shown in Table~\ref{tb:parallax.standard} and
Figure~\ref{fg:parallax.standard.plot}.  Note that all of the measured
CTIOPI parallaxes are within $2\sigma$ of all other observations,
indicating that the current parallax pipeline, DCR corrections, and
conversion from relative to absolute parallax produce reliable
results.  The final parallax error is a combination of many factors,
including (1) the accuracy of the coordinates, (2) the quality of the
reference star frame (brightness, distribution), (3) the accuracy of
the (x,y) centroids, including any ellipticity caused by any close
component (4) the total number of parallax images, (5) the time span
of the available frame series, (6) the parallax factor coverage, (7)
the DCR corrections, and (8) the correction of relative to final
absolute parallax.  The first three factors can not easily be modified
after they are chosen.  However, the number of observations, the
duration of the frames series, and the parallax factor coverage, can
be controlled and depend only on the resources, staffing, and stamina
of the CTIOPI Team.  At present, a pi star is generally considered
``finished'' when all of the following criteria are
met\footnote{Exceptions occur when the pi star is faint, when poor
reference star configurations are available, or when a pi star is
blended with a background source or close physical companion.}:

\begin{enumerate}

\item the relative parallax error is less than 3 mas

\item the pi star has been observed for 4 or more seasons (one season
      includes 2-3 months of observations)

\item the pi star has been observed for at least 2.5 years 

\item there are at least 40 frames of the field

\item $VRI$ photometry has been obtained for the field

\end{enumerate}

In practice, an extended time span results in meeting most of these
criteria, so it is perhaps the best single benchmark to be used to
evaluate parallax errors for the entire survey.
Figure~\ref{fg:pi.time.line} illustrates how time coverage affects the
relative parallax error for 10 different stars within 10 pc (six are
calibration stars and four are additional CTIOPI targets).  Parallax
reductions were executed using various subsets of the complete data
sets (each star indicated with a different symbol).  A few stars show
only $\sim$2 mas error after only about one year of observations.  In
these cases, the parallaxes determined can be quite inaccurate, but a
good fit with minimal formal error can be made to the proper motion
and parallactic motion even though they have not yet been adequately
decoupled.  When key high parallax factor images are taken later, a
different stellar path is determined and the error represents reality.
The mean error for all of the reductions for all 10 fields is 2.45
mas.  This error is reached at a time point 2.32 years into an
observing sequence.  We therefore conclude that $\sim$2.5 years of
coverage is sufficient to determine accurate parallaxes with
acceptable final errors based on the current time baseline we
have. This is consistent with the results of \citet{Dahn2002}, who
find that the USNO parallaxes are stable after about 2 years
observation.

\subsection{Parallax Results for MOTION Stars}

Complete astrometric results for 36 MOTION systems and the seven
calibration stars are presented in Table~\ref{tbl:pi.result}.  These
are the first trigonometric parallaxes for 33 of the MOTION systems
(GJ 545, GJ 754, and LHS 500/501 have improved parallaxes; see
section~\ref{sec:notes} below).  The first two columns are the
identifiers and coordinates.  The third column reports the filter used
for parallax frames.  The next four columns provide observational
statistics.  N$_{sea}$ indicates the number of seasons observed, where
2-3 months of observations count as one season.  The letter ``c''
indicates a continuous set of observations where multiple nights of
data were taken in each season, whereas an ``s'' indicates scattered
observations when some seasons have only a single night of
observations.  Generally, ``c'' observations are better.  A $+$
indicates that three or fewer individual images are used in one or
more seasons that are not counted in N$_{sea}$.  N$_{frm}$ is the
total number of frames used in the final reduction, and Years
indicates the number of years spanned by the full reduction set.
N$_{ref}$ indicates the number of reference stars used during parallax
reductions.  Columns 8-10 report the relative parallax, size of the
relative to absolute parallax correction, and the final absolute
parallax, respectively.  The next two columns are the proper motion
and the direction of proper motion. The thirteenth column is the
derived tangential velocity for each pi star.  The last column has a
``!'', if there are notes in section~\ref{sec:notes}.

\subsection{Notes on Individual Systems}
\label{sec:notes}

Here we comment on individual systems that have ! in the notes column
of Table~\ref{tbl:pi.result}.

{\bf GJ 1050 (LHS 157)} The field lacks bright reference stars, so
some reference stars with fewer than 100 peak counts are included,
causing a relatively large parallax error of 4.44 mas.  The
photometric distance from the \citet{Henry2004} relations is
14.9$\pm$2.2 pc, which is comparable to our trigonometric parallax,
thereby precluding any relatively bright unseen companion that may
cause the high error.

{\bf GJ 1068 (LHS 22)} is a new RECONS sample member at a distance of
6.97 $\pm$ 0.09 pc.  \citet{Ianna1994} reported a preliminary parallax
of 0\farcs1416$\pm$0\farcs0029 (Ianna, 2004, private communication,
not in print).  

{\bf LHS 193AB} is a new multiple system reported in \citet{Jao03}
with a separation of 12\farcs6.  A parallax has been determined only
for LHS 193A because LHS 193B is too faint.  The LHS 193AB system is a
member of the MOTION sample based on the LHS catalog \citep{LHS} value
of $\mu =$ 1\farcs023/yr, but \citet{Bakos2002} flag this object as
having a problematic proper motion.  Our result of $\mu =$
0\farcs9964/yr indicates a proper motion slightly less than
1\arcsec/yr.  We now have a longer time base than given in
\citet{Jao03}, but no orbital motion is detected.  Reference star \#3
(RA $=$ 04 32 25.54, DEC $=$ $-$39 03 14.6, epoch $=$ J2000.0) is
relatively nearby, having $\pi_{rel} =$ 0\farcs03054$\pm$0\farcs00168,
$\mu =$ 0\farcs035/yr, and $V-I = 2.71$, and was dropped from the
final reduction.

{\bf LHS 225AB} is a multiple system reported in \citet{Jao03} and
also in NLTT Catalog \citep{NLTT} with a separation of 2\farcs5.
Parallaxes are determined for both components, but images with
ellipticity greater than 20\% had to be included during data reduction
because of the proximity of the two sources.  This causes both
parallaxes to have relatively high errors.

{\bf GJ 1123 (LHS 263)} is a new RECONS sample member at a distance of
9.02 $\pm$ 0.16 pc. The spectroscopic and photometric distances
estimated by \citet*[7.6pc]{Henry2002} and \citet*[7.5$\pm$1.2
pc]{Henry2004} are both with error less than 17\% from this
measurement.


{\bf GJ 1128 (LHS 271)} is a new RECONS sample member at a distance of
6.53 $\pm$ 0.10 pc, confirming the distance estimates of 6.6 pc in
\citet{Henry2002} and 6.4 $\pm$ 1.0 pc in \citet{Henry2004}. 

{\bf GJ 1129 (LHS 273)} is a new NStars sample member at 11.00 $\pm$
0.46 pc, confirming the distance estimates of 11.6 pc in
\citet{Henry2002}.  Images of the pi star are contaminated by a faint
background star within a few arcseconds throughout the frame series,
and blended during the last two epochs.  This contamination causes the
parallax residuals to have a ''perturbation-like" curve resulting in a
relatively large parallax error of 3.78 mas.  Nonetheless, the
parallax result after the first 2.02 years matches the result after
the full 4.27 years of the current dataset, so the result is reliable.

{\bf DENIS J1048-3956} is a new RECONS sample member at a distance of
4.04 $\pm$ 0.03 pc, confirming the distance estimate of 4.5 $\pm$ 0.7
pc in \citet{Henry2004}.  \citet{Deacon2001} determined a
trigonometric parallax of 0\farcs192$\pm$0\farcs037 using five
SuperCOSMOS photographic plates.  CTIOPI has improved the result to
0\farcs24771$\pm$0\farcs00155\footnote{The result from CTIOPI 1.5m
\citep{Costa2004} is 0\farcs24978$\pm$0\farcs00181}, making DENIS
J1048-3956 the 28th nearest stellar system (after including two stars
that are slightly closer for which we have preliminary, but as yet
unpublished, parallax values).

{\bf LHS 300AB} is a new multiple system reported in \citet{Jao03}
with a separation of 4\farcs3.  A mixture of resolved and unresolved
images are included in the dataset, but because the B component is 4.9
mag fainter than A in the filter chosen for parallax frames, $R$, the
centroid is not significantly corrupted by B.

{\bf LHS 382} is close to the ecliptic, so the axis of the parallactic
ellipse is small in the Y direction. Strong nebulosity is seen in this
field.

{\bf LTT 6933 (LHS 3292)} is a member of the MOTION sample based on
the \cite{Bakos2002} value of $\mu =$ 1\farcs03/yr derived using
POSS-I and POSS-II plates separated by 13.8 years.  The LHS catalog
reports $\mu =$ 0\farcs996/yr, which is confirmed by our result of
$\mu =$ 0\farcs9593/yr.

{\bf GJ 1226AB (LHS 263AB)} is a multiple system reported in
\citet{Jao03} and \citet{Vilkki1984} with a separation of 1\farcs4.
Parallaxes were determined for both components.  All 105 available
frames were examined manually and only 59 images with good seeing were
selected for data reduction.  The absolute parallax correction for
this field is over 5 mas and it is much larger than our mean
corrections. Consequently, the mean correction 1.47 mas is
adopted. Further investigation is necessary.

The Yale Parallax Catalog \citep{YPC} gives parallaxes for {\bf GJ 545
(LHS 369)}, $\pi_{trig}$ $=$ 0\farcs0911$\pm$0\farcs015, {\bf GJ 754
(LHS 60)}, $\pi_{trig}$ $=$ 0\farcs1752$\pm$0\farcs0101, and {\bf LHS
500/501}, $\pi_{trig}$ $=$ 0\farcs075$\pm$0\farcs0171).  Our results
have significantly improved the parallaxes by factors of 11, 7, and
11, respectively.  LHS 500/501 is a wide binary with separation
107\arcsec~for which we have determined parallaxes for both
components.  The two parallaxes are entirely consistent, differing by
1.8$\sigma$.

{\bf Proxima Centauri (LHS 49)} is one of our parallax calibration
stars.  Proxima is brighter than the 0.9-m telescope limit in the $I$
band, so $VRI$ from \citet{Bessel1990} has been adopted for it, with
proper transformations to the Johnson-Kron-Cousins system used for the
calculation of the DCR corrections.  The last epoch of data presented
here (Dec 2003) was taken at an hour angle greater than 4 hours, so
Stone's (1996) theoretical model was used for DCR, rather than the
empirical model.  We note that our value of $\pi_{trig}$ $=$
0\farcs77425$\pm$0\farcs00208 is the most precise ground-based
parallax ever determined for the nearest star to our Solar System, and
has a formal error 13\% smaller than the Hipparcos result.

\section{$VRIJHK_{s}$ Photometry Results}

The $VRIJHK_{s}$ photometry for the 48 stars in 43 systems is
presented in Table~\ref{tbl:phot.result}.  After the two names, the
next four columns are the new optical $VRI$ photometry and the number
of new observations taken during CTIOPI.  For comparison purposes,
references for previously published photometry are listed in the
seventh column.  The next three columns are the infrared $JHK_{s}$
photometry (rounded to the hundredth) from 2MASS.  Spectral types and
references are given in the last two columns.

The $VRI$ data have been reduced as discussed in
section~\ref{sec:phot.reduce}.  Most of the stars are reduced using an
aperture 7\arcsec~in radius.  A few stars required smaller aperture
sizes in order to separate two close components: LHS 193B (4\arcsec),
LHS 225AB (2\arcsec), LHS 300AB (2\arcsec), and GJ 1226AB (1\arcsec).
Errors from the fits of standard stars (external errors) are estimated
to be $\pm$0.02 at $V$, $R$ and $I$.  Because most of the pi stars are
bright, the signal to noise ratio errors (internal errors) are usually
from 0.001 to 0.008 mag.  The exceptions are LHS 193B (0.04, 0.05,
0.05 mag at $VRI$, respectively), DENIS 1048-3956 (0.02 at $V$ band),
and LHS 300B (0.01, 0.02, 0.02).  We estimate that night-to-night
repeatability errors for the faintest stars in the CTIOPI program (the
worst case) are $\sim$0.03, as discussed in \citet{Henry2004}, except
for those stars that are possibly variable, e.g.~DEN 1048-3956.  Thus,
the combination of all three errors for the relatively bright stars
presented here is typically $\sim$0.03 mag at $VRI$.

Infrared photometry in the $JHK_{s}$ system has been extracted from
2MASS.  The $JHK_{s}$ magnitude errors from the total photometric
uncertainties, including global and systematic terms, are almost
always less than 0.05 mag and are typically 0.02-0.03 mag. The
exceptions are LHS 193B (errors of 0.11, 0.16 and 0.18 at $JHK_{s}$,
respectively), LHS 271 (0.05 at $H$), Proxima Cen (0.06 at $H$) and
LHS 3292 (0.06 at $H$).

\section{Discussion}
\label{discussion}

In this paper, the CTIOPI team presents the first substantial set of
trigonometric parallaxes for stars with proper motion greater than
1\farcs0/year since the Yale Parallax Catalog and the Hipparcos
mission.  \citet{Hambly1999}, \citet{Dahn2002}, \citet{Tinney2003} and
\citet{Vrba2004} have reported a total of 1, 1, 2 and 5 first
trigonometric parallaxes for MOTION systems, respectively.  All of
those studies concentrated on the (very) cool end of main sequence, L
or T dwarfs, or in a single case, a white dwarf.  Obviously, the
MOTION systems are potentially nearby stars, and this is borne out by
our results --- Table~\ref{tbl:pi.result} shows that four of the
systems are new entrants to the RECONS 10 pc sample, which requires a
reliable trigonometric parallax published in a refereed journal for
inclusion.  Furthermore, 22 additional systems are new members of the
NStars (25 pc) sample, and only seven systems lie beyond the NStars
horizon.  In sum, the first trigonometric parallaxes reported here for
33 MOTION systems provide reliable distances to 41\% of the MOTION
systems south of $\delta$ $=$ 0 that previously had no trigonometric
distance measurements.

The combination of accurate $\pi_{trig}$ and $VRIJHK_{s}$ photometry
permits the construction of reliable HR diagrams and offers the
opportunity for insight into the MOTION sample.  Here we present HR
diagrams for the MOTION stars with new and improved parallaxes from
CTIOPI, split into single-star systems and binaries (the parallax
standard stars are not included in this discussion).  In particular,
we discuss the identification of new nearby subdwarfs, and two
remarkable new K/M type subdwarf-white dwarf binaries (hereafter,
sdK/M+WD).

\subsection{HR Diagram for Single MOTION Stars}

In Figure~\ref{fg:color.mag.1}, we plot $M_{K_{s}}$ against the $V-K_{s}$
color for all stars in Table~\ref{tbl:phot.result}, excluding the
parallax calibration stars and the five binary systems in the sample
--- LHS 193AB, LHS 225AB, LHS 300AB, LHS 500/501, and GJ 1226AB.
These binaries will be discussed in the next section.  Because of the
high quality $\pi_{trig}$ and $VRIJHK_{s}$ photometry, the errors in
$M_{K_{s}}$ and $V-K_{s}$ are roughly the size of the symbols.

By comparing our sample with the main sequence stars from the RECONS
10 pc sample and subdwarfs from \citet{Gizis1997}, we can estimate the
luminosity classes for several stars without spectral types.  Of the
31 single MOTION stars, the seven labeled on the plot (and indicated
with open circles) do not have spectral types.  Three of these are new
subdwarfs --- LHS 158 at 40.1 pc, LHS 382 at 48.3 pc and LHS 521 at
46.3 pc.  It is clear from Figure~\ref{fg:color.mag.1} that LHS 521 is
an extreme subdwarf.  The remaining four stars without spectral types
are main sequence stars that are all new NStars members --- ER 2 at
11.9 pc, WT 1827 at 12.3 pc, pc LTT 6933 at 16.4 pc, and LHS 539 at
18.9 pc.  Among the 24 stars with spectral types, three have been
misclassified as main sequence stars, but are likely to be nearby
subdwarfs --- LHS 406 at 21.1 pc ($M_{K_{s}}=$ 7.39, $V-K_{s}=$ 4.04), WT
248 at 26.0 pc ($M_{K_{s}}=$ 7.79, $V-K_{s}=$ 4.65), and LHS 440 at 27.1
pc ($M_{K_{s}}=$ 6.79, $V-K_{s}=$ 4.03).  Spectroscopic observations are
necessary to confirm their luminosity classes.

This highly kinematically biased sample is of course likely to include
Galactic thick disk members and even a few high velocity field halo
subdwarfs.  The tangential velocities of the new subdwarfs LHS 158
(191 km/sec), LHS 382 (327 km/sec), and LHS 521 (221 km/sec) are,
indeed, quite high, implying that they belong to an old population.
In order to analyze the full kinematics for these systems, future
radial velocity observations are necessary.

\subsection{HR Diagram for Binary MOTION Stars}
\label{sec:binary}

Binary systems provide several opportunities to glean additional
insight into stellar properties because the components are assumed to
have formed simultaneously (so have the same age), and from the same
gas cloud (so have identical composition).  If parallaxes can be
determined for both stars in a binary, a consistent match also
indicates that our observing and reduction methodology is sound.  The
five binary systems in Tables 2 and 3 are shown on the $M_{V}$
vs.~$V-I$ HR diagram in Figure~\ref{fg:color.mag.2}. $M_{V}$ has been
used instead of $M_{K_{s}}$ because $K_{s}$ magnitudes are not available
for the individual components in LHS 225AB, LHS 300AB, and GJ 1226AB.
As discussed in the previous section, the RECONS and subdwarf samples
are plotted for comparison, and have been supplemented in
Figure~\ref{fg:color.mag.2} with white dwarfs from \citet[hereafter,
BLR]{Bergeron2001}.

It is clear from Figure~\ref{fg:color.mag.2} that the components of
the close binaries LHS 225AB and GJ 1226AB are nearly identical main
sequence M dwarfs.  Our separate $\pi_{trig}$ determinations for the
components of the wide LHS 500/501 pair are consistent, and show that
the two components are both main sequence M dwarfs.

The two remaining binaries, LHS 193AB and LHS 300AB, are both
comprised of a subdwarf of late K/early M type and a white dwarf.  A
search of the literature indicates that few such systems are known.
\citet{Gizis1997b} reported that LHS 2139/2140 is a common proper
motion sdK/M+WD pair, based on a noisy but featureless spectrum for
the B component, but no available parallaxes are available to confirm
the nature of the system.  \citet{Gizis1998} argued that GJ 781AB, an
unresolved spectroscopy binary, is another sdK/M+WD binary based on
its mass function.

The two new wide sdK/M+ WD pairs reported here have complete parallax
and $VRI$ photometry.  In both cases, Figure~\ref{fg:color.mag.2}
clearly indicates that the primary star is a subdwarf and that the
secondary is a white dwarf.  However, both white dwarfs are redder
than all white dwarfs with $V-I$ and parallax reported in BLR. The
locations of LHS 193B and LHS 300B in the HR diagram can possibly be
explained several ways, including multiplicity, composition, very low
mass (and hence large size), and/or dust.

In particular, models by \citet{Bergeron1995} indicate that very low
mass helium white dwarfs may have the colors observed.  Both
\citet{Hansen1998} and \citet{Bergeron2001b} argue that as the
$T_{eff}$ decreases to 3000 K for old ($t \gtrsim$ 11 Gyr) white
dwarfs with hydrogen atmospheres, their location in the HR diagram
swings back $blueward$ of the white dwarf cooling sequence. This is
caused by strong H$_{2}$ molecular absorption features expanding into
the optical regions.  This implies that both LHS 193B and LHS 300B,
which lie outside the grid for typical hydrogen white dwarfs, may be
helium white dwarfs (or hydrogen white dwarfs with lower surface
gravity values than have been included in the model grid -- unlikely,
because of the observed distribution of surface gravities for hydrogen
white dwarfs).  Very low S/N spectra currently in hand indicate that
the two white dwarfs are featureless, in particular having no
H$\alpha$ line.  Obviously, high S/N spectra are desirable, and will
be the focus of future work.

\section{Conclusions}

Accurate $\pi_{trig}$ and $VRIJHK_{s}$ for nearby stars assist in
constructing the basic framework of stellar astronomy.  Here we
provide a valuable contribution to studies of the solar neighborhood
by targeting MOTION stars. A total of 46 parallaxes from CTIOPI are
presented, including 39 parallaxes for 36 MOTION systems and 7
additional parallaxes for calibration stars.  Thirty-three MOTION
systems have trigonometric parallaxes determined for the first time.

Already, several new nearby systems have been revealed.  Four of the
MOTION systems --- GJ 1068, GJ 1123, GJ 1128 and DENIS J1048-3956 ---
are new members of the RECONS 10 pc sample \citep{Henry1997}.  An
additional 22 systems are new members of the NStars 25 pc sample
\citep{Henry2003}.  In addition, valuable new nearby subdwarfs have
been identified, and two rare sdK/M+WD pairs have been discovered.
Both of these samples are valuable probes of the history of our
Galaxy.

This work once again shows that faint, high proper motion stars are
excellent candidates to discover nearby stars.  Yet, 48 MOTION systems
south of $\delta$ $=$ 0 still do not have parallaxes.  In future
papers, we will present $\pi_{trig}$ and $VRIJHK_{s}$ for several
additional samples of stars, including more MOTION systems, stars
neglected in the LHS Catalog, and new discoveries from our SuperCOSMOS
RECONS search (\citealt*{Hambly2004}, \citealt*{Henry2004},
\citealt*{Subasavage2005}), as well as others.

Finally, CTIOPI has been expanded in recent years under the SMARTS
Consortium to carry out a program called ASPENS (Astrometric Search
for Planets Encircling Nearby Stars), led by David Koerner at Northern
Arizona University.  Red and white dwarf systems within 10 pc south of
$\delta$ $=$ 0, including the four new RECONS members and six of the
seven calibration stars, are being observed intensely to reveal any
possible long term astrometric perturbations.

\section{Acknowledgments}

We would like to thank Barbara McArthur, Mike Begam, Dave Monet, and
Myles Standish for their assistance in building the data reduction
pipeline.  We gratefully acknowledge assistance in the early stages of
the CTIOPI effort from Claudio Anguita, Rafael Pujals, Maria Teresa
Ruiz and Pat Seitzer. Without the extensive observing support of
Alberto Miranda, Edgardo Cosgrove, Arturo Gomez and the staff at CTIO,
CTIOPI would not be possible.  We also thank Jacob Bean, Thom
Beaulieu, Charlie Finch, and Jennifer Winters for their assistance on
data organization, reduction and observing.  We thank Pierre Bergeron,
John Gizis, Hugh Harris, Dave Latham, James Liebert, Hektor Monteiro,
and Terry Oswalt for suggestions concerning the subdwarf-white dwarf
binaries.

We are deeply indebted to NOAO for providing us a long term observing
program at CTIO via the NOAO Surveys Project, using both the CTIO
0.9-m and 1.5-m telescopes.  We also thank the continuing support of
the members of the SMARTS Consortium without whom the completion of
many astrometric series reported here would not have been
possible. The early phase of CTIOPI was supported by the NASA/NSF
Nearby Star (NStars) Project through NASA Ames Research Center. The
RECONS team at Georgia State University is supported by NASA's Space
Interferometry Mission and GSU.  This work has used data products from
the Two Micron All Sky Survey, which is a joint project of the
University of Massachusetts and the Infrared Processing and Analysis
Center at California Institute of Technology funded by NASA and NSF.

EC and RAM acknowledge support by the Fondo Nacional de Investigaci\'on
Cient\'ifica y Tecnol\'ogica (proyecto Fondecyt No. 1010137), and by the
Chilean Centro de Astrof\'isica  FONDAP (No. 15010003).
This project has made generous use of the 10\% Chilean time.


\clearpage


\figcaption[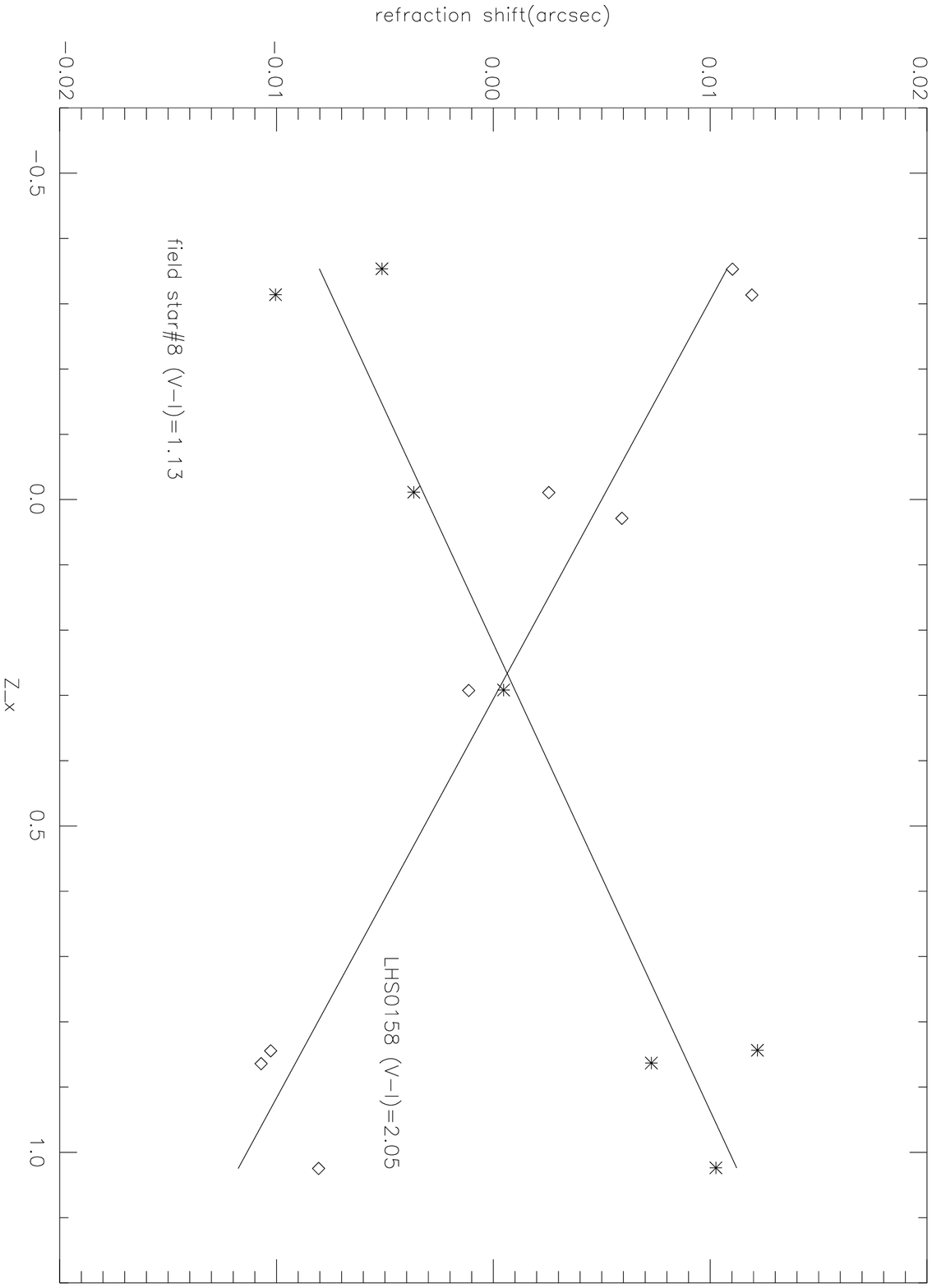]{This plot illustrates examples of atmospheric
refraction shift for two stars having different colors.  The slope for
LHS 158 is $-$0.016, while the slope for field star \#8 is $+$0.014.
There is no point for field star \#8 at $Z_{x}\approx$0.02 because of
poor image quality.
\label{fig:dcr.slope}}

\figcaption[dcr.fit.plot]{These three plots illustrate the DCR data
taken at the CTIO 0.9-m for ten parallax fields and the empirical least
squares fits (solid lines) used in the final astrometric reductions
for all three filters used in CTIOPI.  The top panel is for the $V$
filter, for which a second-order fit is used.  The middle panel is for
the $R$ filter, for which a fourth-order fit is used.  The bottom
panel is for the $I$ filter, for which a first-order fit is used.  The
curves from the theoretical models of \citet{Stone1996} are plotted as
dashed lines.
\label{fig:dcr.fit.plot}}

\figcaption[gj1061.dcr.plot]{Effects of DCR corrections in the GJ 1061
field are illustrated for both the X and Y directions. These four
plots indicate the X and Y residuals with (right two plots) and
without (left two plots) DCR correction. The only images with HA
greater than 100 mins are at the last epoch. The improvement in the X
and Y residuals after DCR corrections is evident in the two plots in
the right, especially in the X direction.
\label{fg:dcr.gj1061}}

\figcaption[pi.standard.comparison]{Comparison of CTIOPI parallaxes to
parallaxes measured from the ground (YPC $=$ Yale Parallax Catalog),
and space (HIP $=$ Hipparcos, HST $=$ Hubble Space Telescope).  Note
that the CTIOPI parallaxes generally have errors smaller than those
from YPC and comparable to those from HIP.  The units on the Y axis
are arcseconds.  \label{fg:parallax.standard.plot}}

\figcaption[time.coverage]{This plot shows the relation between
relative parallax error and time coverage for 10 different nearby
stars (each plotted using a different symbol).  The dashed line
indicates the mean error for all reductions of the 10 fields.  The
crossover point of the fit to the decreasing errors and the mean error
for all reductions is at 2.32 years.
\label{fg:pi.time.line}}

\figcaption[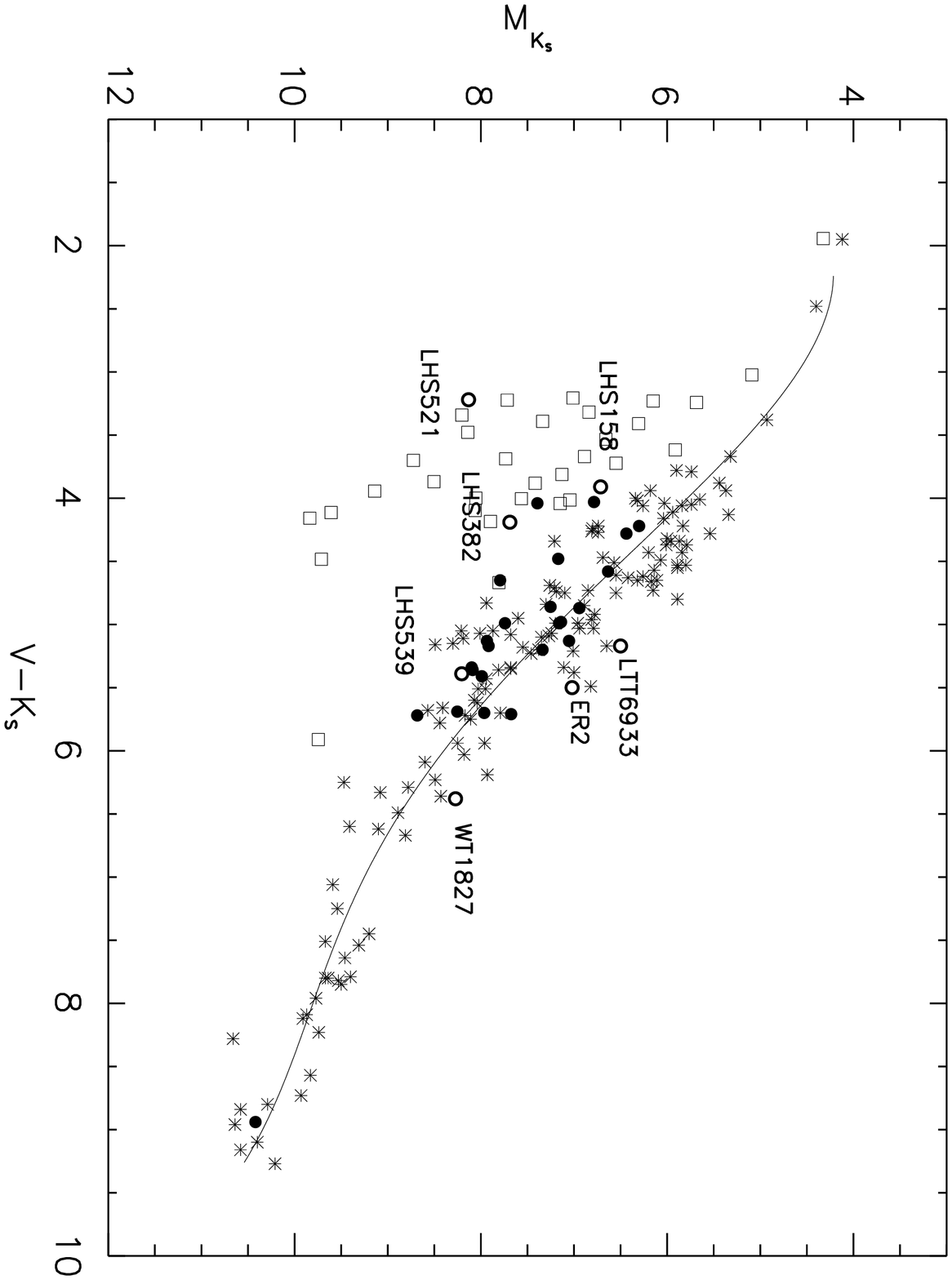]{This HR diagram, using $M_{K_{s}}$
vs.~$V-K_{s}$, is shown for single MOTION stars with new and improved
parallaxes.  Filled circles represent 24 MOTION stars with published
spectral types given in Table~\ref{tbl:phot.result}.  Open circles
represent 7 MOTION stars without spectral types.  Open boxes represent
32 subdwarfs (LHS stars with $\mu >~$1\farcs0/yr) from
\citet{Gizis1997}.  Asterisks represent RECONS sample members and some
very late M dwarfs discussed in \citet{Henry2004}, with an empirical
fit tracing the main sequence stars.  The single solid point at the
far right is DENIS J1048-3956.
\label{fg:color.mag.1}}

\clearpage

\figcaption[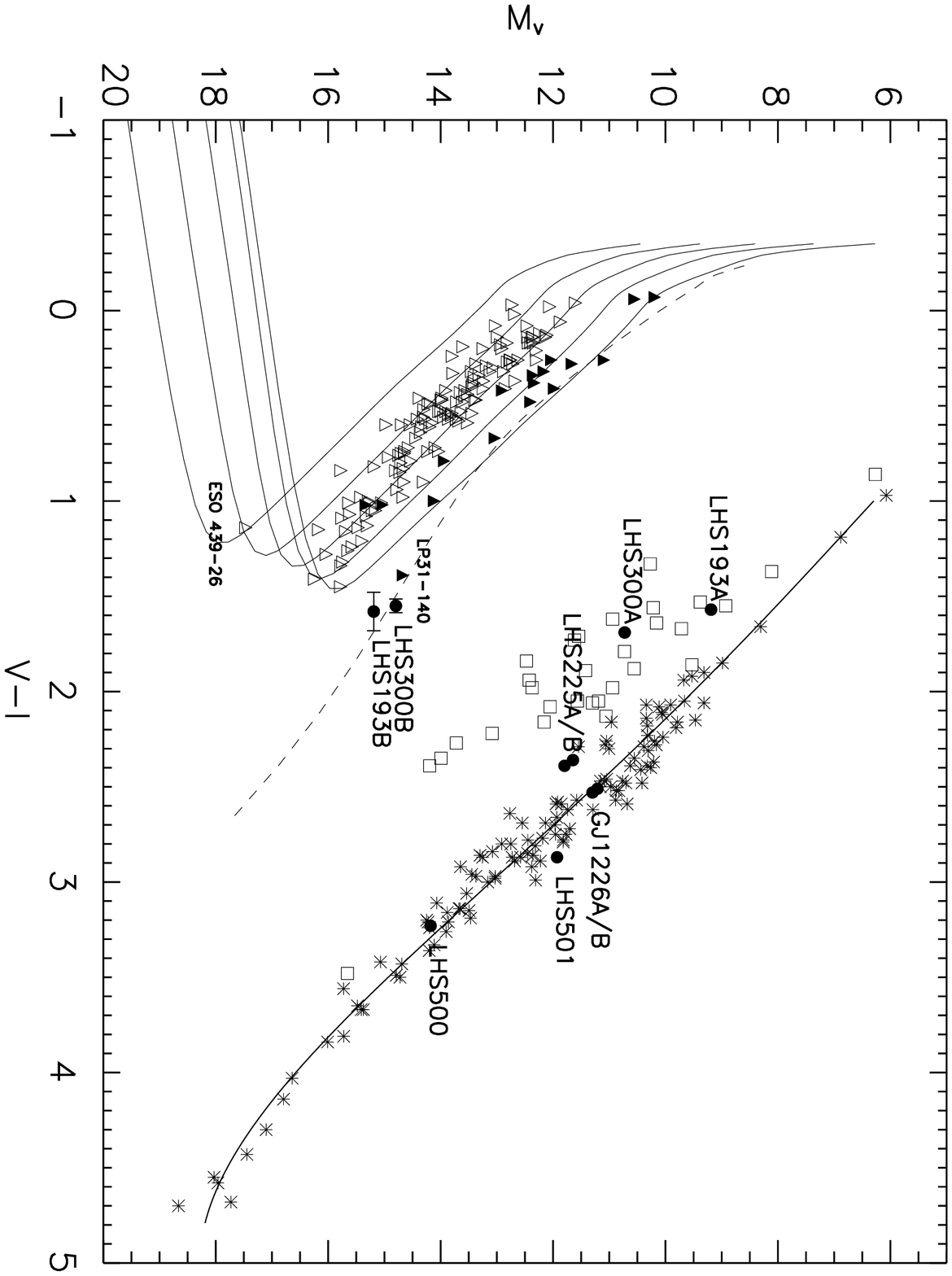]{This HR diagram, using $M_V$ vs.~$V-I$, is
shown for all components (solid points) in the five binary MOTION
systems with new and improved parallaxes discussed in
section~\ref{sec:binary}.  Symbols are the same as defined in
Figure~\ref{fg:color.mag.1}, supplemented with additional points for
white dwarfs from BLR (open triangles are single white dwarfs, filled
triangles are known or suspected double degenerates).  The grid of
models from \citet{Bergeron1995} outlines the extent of the hydrogen
white dwarf sequence.  The dotted line is for pure helium white dwarfs
with {\it log g = 7}, also from \citet{Bergeron1995}.  ESO 439-26 is a
massive (1.19 M$_{\sun}$), pure helium atmosphere white dwarf (BLR).
LP 31-140 is a very low mass (0.19 M$_{\sun}$) and low T$_{eff}$ (4650
K) white dwarf (BLR).  The error bars shown for the $V-I$ colors of
LHS 193B and LHS 300B are from the Poisson errors, whereas the errors
in $M_{V}$ are smaller than the filled circles.\label{fg:color.mag.2}}


\begin{figure}
\includegraphics[angle=90,scale=0.7]{jao.fig1.ps}
\end{figure}

\begin{figure}
\begin{center}
\includegraphics[scale=0.45,angle=90]{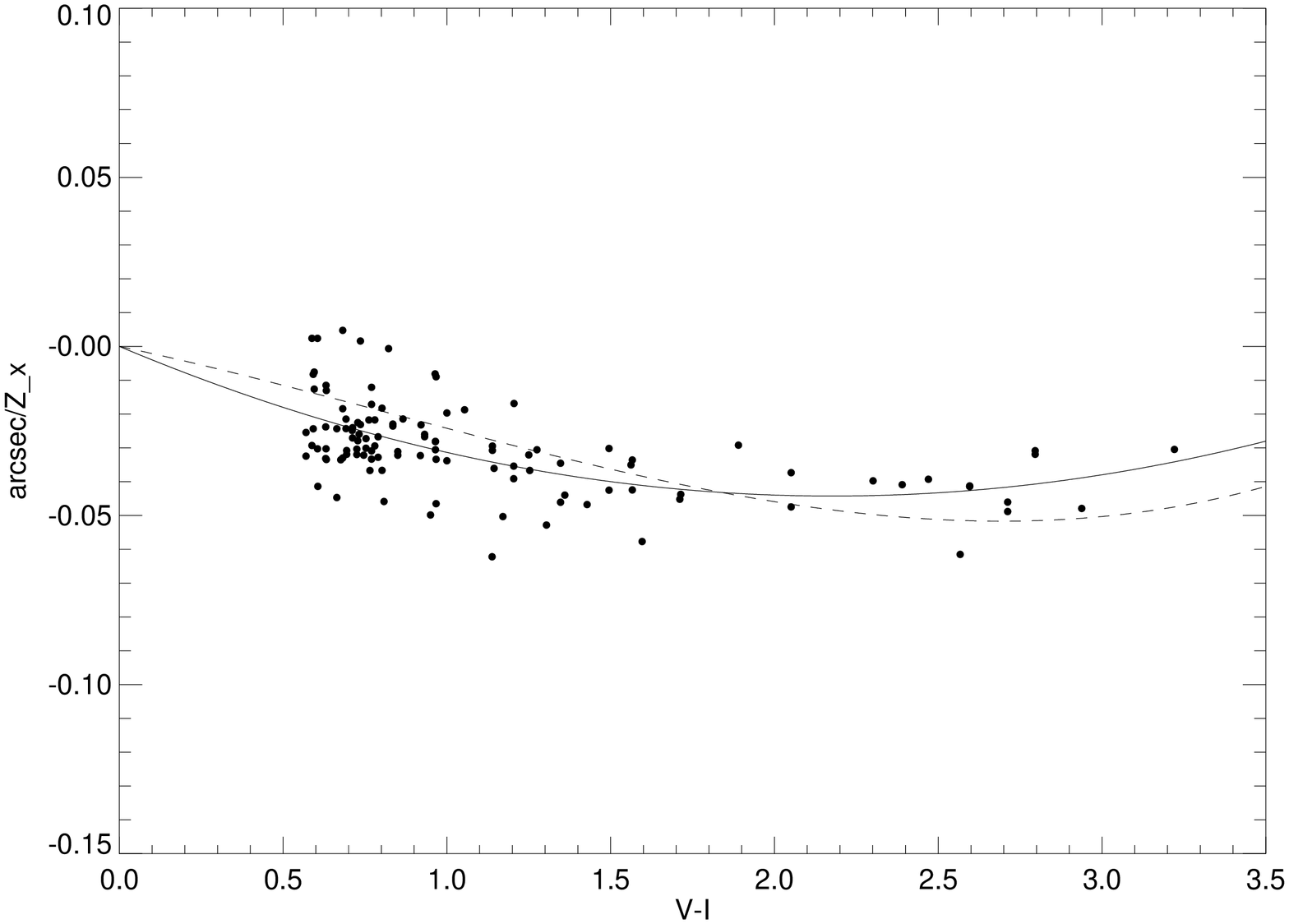}\\
\includegraphics[scale=0.45,angle=90]{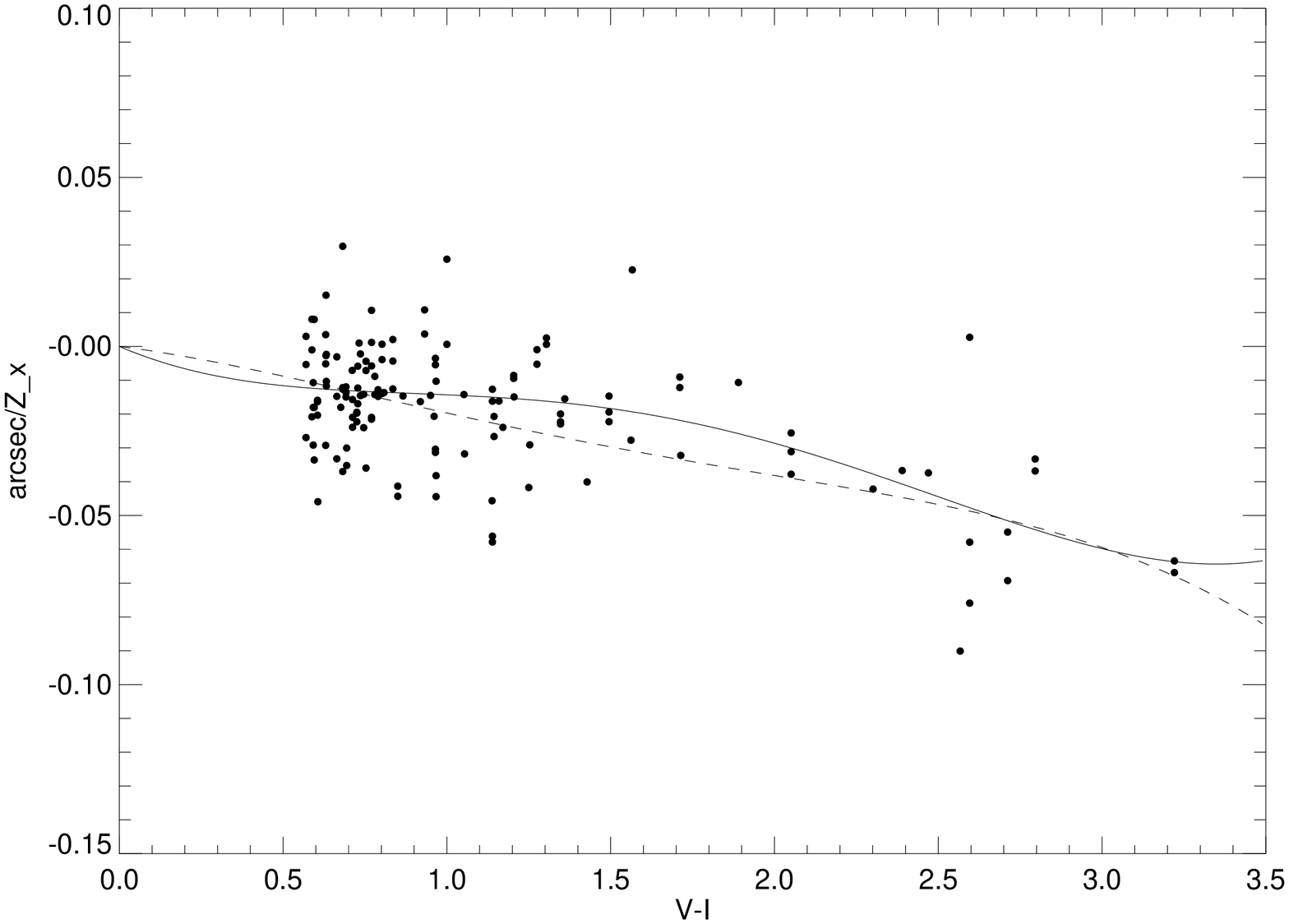}\\
\includegraphics[scale=0.45,angle=90]{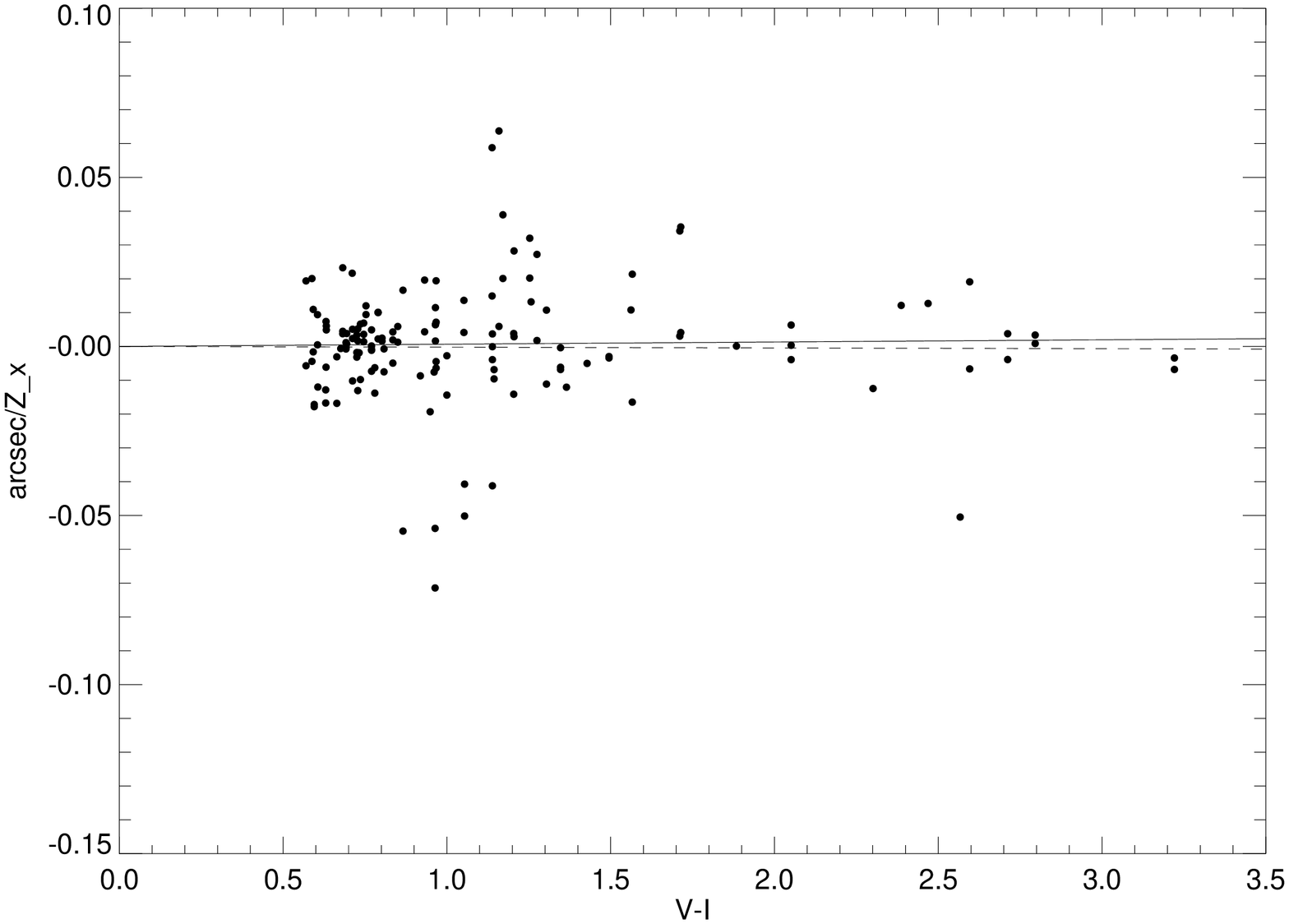}
\end{center}
\end{figure}

\begin{figure}
\includegraphics[angle=180,scale=0.8]{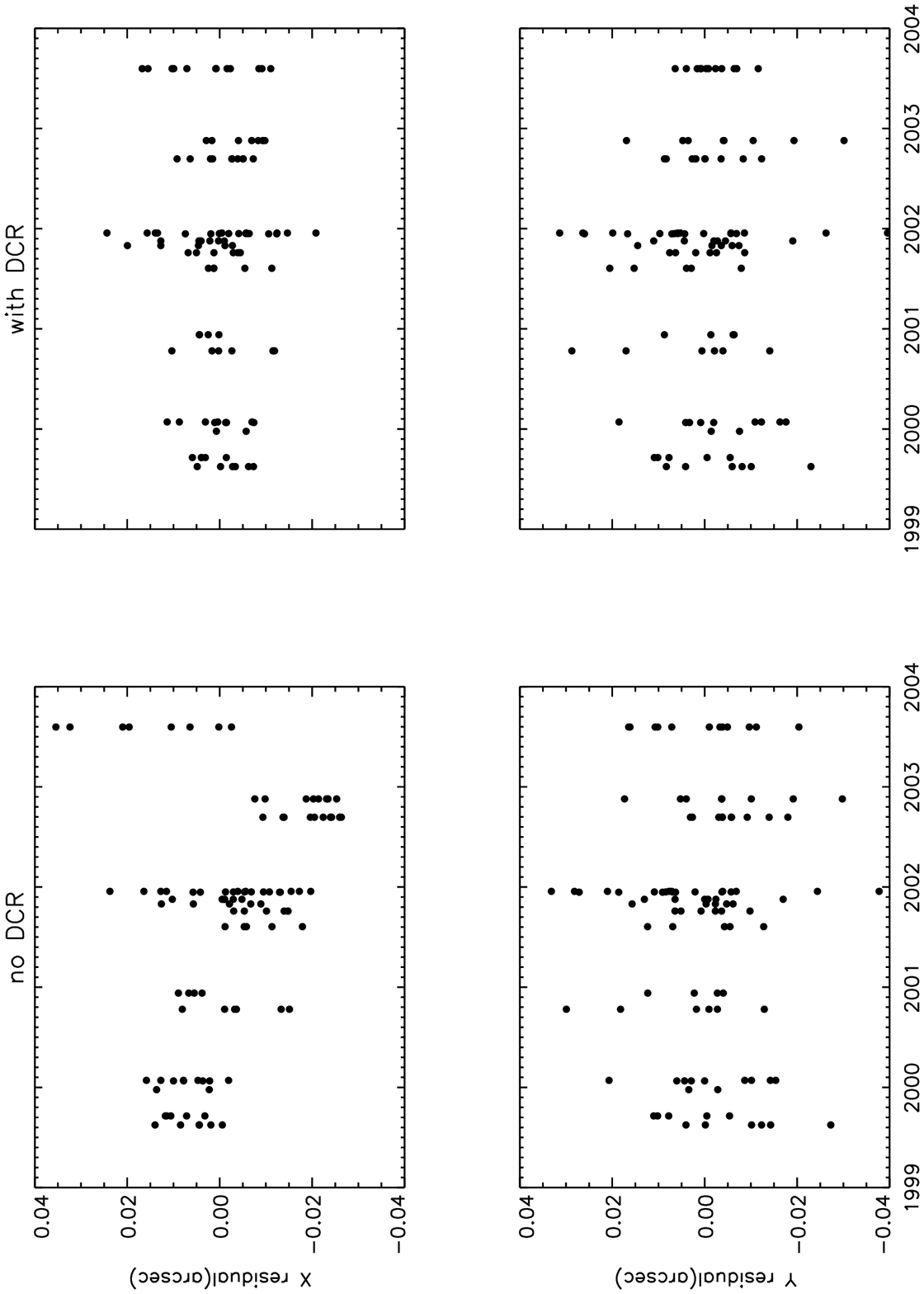}
\end{figure}

\begin{figure}
\includegraphics[angle=90,scale=0.7]{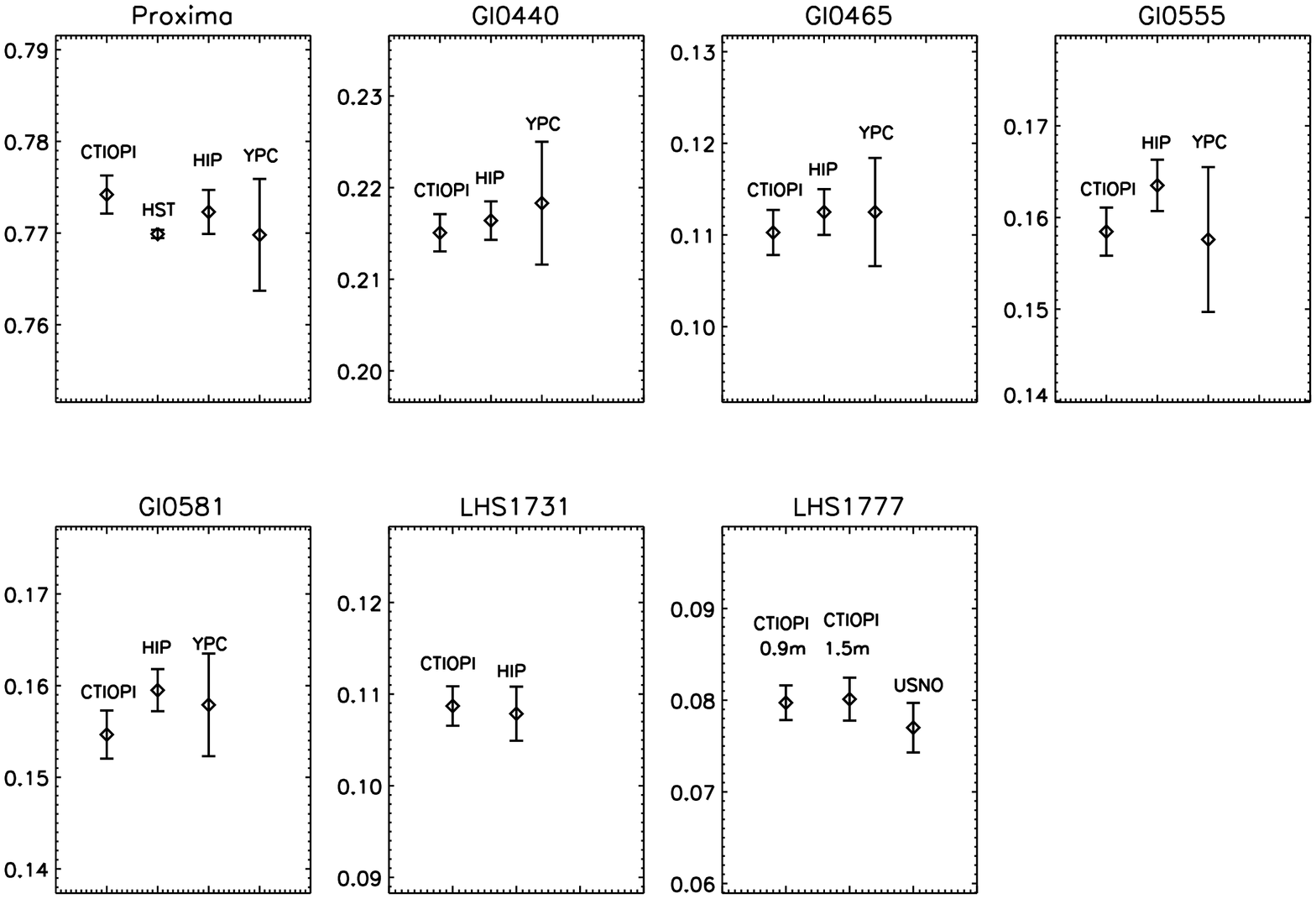}
\end{figure}

\begin{figure}
\includegraphics[angle=90,scale=0.7]{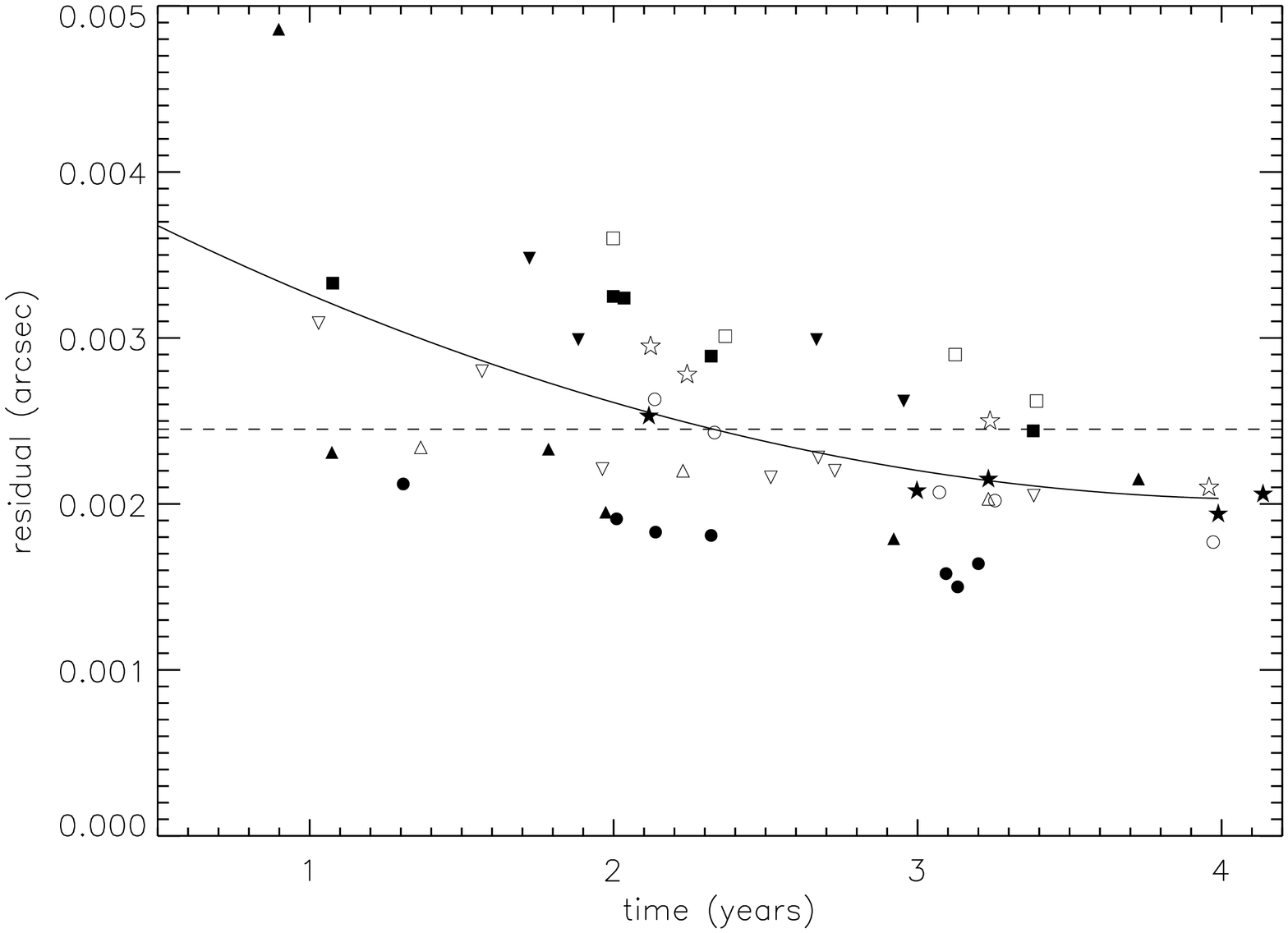}
\end{figure}

\begin{figure}
\includegraphics[angle=90, scale=0.7]{jao.fig8.ps}
\end{figure}

\begin{figure}
\includegraphics[angle=90, scale=0.7]{jao.fig9.ps}
\end{figure}

\clearpage


\begin{table}
\centering
\caption{The parallax standard stars from CTIOPI.}
\label{tb:parallax.standard}
\footnotesize
\begin{tabular}{ccccc}
\tableline\tableline
                           & CTIOPI              & Hipparcos           & YPC                 & others                               \\
\tableline
Proxima\tablenotemark{a}   & 0.77425$\pm$0.00208 & 0.77230$\pm$0.00240 & 0.76980$\pm$0.00610 & 0.76991$\pm$0.00054\tablenotemark{b} \\
GJ 440\tablenotemark{a}    & 0.21507$\pm$0.00204 & 0.21640$\pm$0.00210 & 0.21830$\pm$0.00670 & \nodata                              \\
GJ 465\tablenotemark{a}    & 0.11027$\pm$0.00246 & 0.11250$\pm$0.00250 & 0.11250$\pm$0.00590 & \nodata                              \\
GJ 555                     & 0.15846$\pm$0.00262 & 0.16350$\pm$0.00280 & 0.15760$\pm$0.00790 & \nodata                              \\
GJ 581\tablenotemark{a}    & 0.15466$\pm$0.00262 & 0.15950$\pm$0.00230 & 0.15790$\pm$0.00560 & \nodata                              \\
LHS 1731                   & 0.10870$\pm$0.00215 & 0.10786$\pm$0.00295 & \nodata             & \nodata                              \\
LHS 1777                   & 0.07972$\pm$0.00189 & \nodata             & 0.07820$\pm$0.00270 & 0.08011$\pm$0.00234\tablenotemark{c} \\
\tableline
\multicolumn{5}{c}{}                                                                                                                \\
\end{tabular}

\tablecomments{The numbers indicate $\pi\pm$error in arcseconds.}
\tablenotetext{a}{The star is a member of the MOTION sample.}
\tablenotetext{b}{The parallax error from HST is slightly different
from \citet{Benedict1999} because of improvement in the calculation of
the correction from relative to absolute parallax (Benedict, 2003
private communication).}  \tablenotetext{c}{The result is from CTIOPI
1.5-m \citep{Costa2004}}
\end{table}


\begin{deluxetable}{lcccccccrrrrrrc}
\rotate
\setlength{\tabcolsep}{0.03in}
\tablewidth{0pt}
\tabletypesize{\scriptsize}
\tablehead{\colhead{Name}                &
	   \colhead{RA}                  &
 	   \colhead{DEC}                 &
 	   \colhead{Filt}                &
	   \colhead{Nsea}                &
	   \colhead{Nfrm}                &
	   \colhead{Years}               &
	   \colhead{Nref}                &
	   \colhead{$\pi$(rel)}          &
	   \colhead{$\pi$(corr)}         &
	   \colhead{$\pi$(abs)}          &
	   \colhead{$\mu$}               &
	   \colhead{P.A.}                &
	   \colhead{$V_{tan}$}           &
	   \colhead{Note}                \\
	   \colhead{}                    & 
	   \multicolumn{2}{c}{(J2000.0)} &
           \colhead{}                    &
	   \colhead{}                    &
	   \colhead{}                    &
	   \colhead{}                    &
	   \colhead{}                    &
	   \colhead{(mas)}               &
	   \colhead{(mas)}               &
	   \colhead{(mas)}               &
	   \colhead{(mas/yr)}            &
	   \colhead{(deg)}               &
	   \colhead{(km/s)}              &
	   \colhead{}                    \\
           \colhead{(1)}                 &
           \multicolumn{2}{c}{(2)}       &
           \colhead{(3)}                 &
           \colhead{(4)}                 &
           \colhead{(5)}                 &
           \colhead{(6)}                 &
           \colhead{(7)}                 &
           \colhead{(8)}                 &
           \colhead{(9)}                 &
           \colhead{(10)}                &
           \colhead{(11)}                &
           \colhead{(12)}                &
           \colhead{(13)}                &
           \colhead{(14)}                }
\startdata
\multicolumn{15}{c}{First Trigonometric Parallaxes}\\				   
\hline
GJ 1025              &   01 00 56.37 & $-$04 26 56.5 &       V  &     4c &  46 &  3.28 &   7  &     87.13$\pm$2.38  & 0.57$\pm$0.05  &  87.70$\pm$2.38  &  1311.7$\pm$1.7  &  70.7$\pm$0.13  &  70.9  &       \\
GJ 1050              &   02 39 50.71 & $-$34 07 57.5 &       V  &     4c &  71 &  3.21 &   5  &     92.33$\pm$4.42  & 1.41$\pm$0.44  &  93.74$\pm$4.44  &  1736.9$\pm$3.9  & 162.5$\pm$0.23  &  87.8  &     ! \\
LHS 158              &   02 42 02.88 & $-$44 30 58.7 &       I  &     4c &  84 &  2.91 &  10  &     23.23$\pm$1.61  & 1.71$\pm$0.12  &  24.94$\pm$1.61  &  1005.2$\pm$1.3  &  87.6$\pm$0.13  & 191.0  &       \\
GJ 1068              &   04 10 28.14 & $-$53 36 08.2 &       R  &     5c &  76 &  3.96 &   8  &    142.39$\pm$1.92  & 1.03$\pm$0.12  & 143.42$\pm$1.92  &  2553.5$\pm$1.6  & 199.1$\pm$0.06  &  84.4  &     ! \\
LHS 193A             &   04 32 36.56 & $-$39 02 03.4 &       V  &     4c &  73 &  2.90 &   8  &     29.85$\pm$1.64  & 2.21$\pm$0.21  &  32.06$\pm$1.65  &   996.4$\pm$2.4  &  44.8$\pm$0.28  & 147.3  &     ! \\
LHS 225B             &   07 04 45.77 & $-$38 36 07.6 &       V  &     4c &  74 &  3.29 &   9  &     52.10$\pm$3.83  & 1.96$\pm$0.10  &  54.06$\pm$3.83  &  1211.6$\pm$3.5  & 102.7$\pm$0.28  & 106.2  &     ! \\
LHS 225A             &   07 04 45.77 & $-$38 36 07.6 &       V  &     4c &  74 &  3.29 &   9  &     54.95$\pm$2.67  & 1.96$\pm$0.10  &  56.91$\pm$2.67  &  1191.8$\pm$2.4  & 102.6$\pm$0.20  &  99.3  &     ! \\
GJ 1118              &   08 59 05.31 & $-$31 13 26.6 &       R  &     4c &  60 &  3.18 &  11  &     54.78$\pm$1.76  & 1.41$\pm$0.11  &  56.19$\pm$1.76  &  1094.0$\pm$1.5  & 139.9$\pm$0.15  &  92.3  &       \\
GJ 1123              &   09 17 05.33 & $-$77 49 23.4 &       V  &     4s &  40 &  4.10 &  10  &    108.34$\pm$2.01  & 2.58$\pm$0.17  & 110.92$\pm$2.02  &  1051.7$\pm$2.1  & 141.6$\pm$0.23  &  44.9  &     ! \\
GJ 1128              &   09 42 46.36 & $-$68 53 06.0 &       V  &     3c+&  71 &  4.10 &  11  &    151.45$\pm$2.40  & 1.60$\pm$0.24  & 153.05$\pm$2.41  &  1120.5$\pm$2.5  & 365.7$\pm$0.19  &  34.7  &     ! \\
GJ 1129              &   09 44 47.34 & $-$18 12 48.9 &       V  &     4s &  55 &  4.27 &   8  &     89.23$\pm$3.78  & 1.70$\pm$0.13  &  90.93$\pm$3.78  &  1590.0$\pm$1.7  & 265.2$\pm$0.09  &  82.9  &     ! \\
WT 248               &   10 05 54.94 & $-$67 21 31.2 &       I  &     4c &  52 &  3.95 &  11  &     37.32$\pm$2.83  & 1.12$\pm$0.08  &  38.44$\pm$2.83  &  1211.6$\pm$2.0  & 265.2$\pm$0.13  & 149.4  &       \\
LHS 281              &   10 14 51.77 & $-$47 09 24.1 &       R  &     4c &  60 &  3.19 &  10  &     82.45$\pm$1.69  & 0.62$\pm$0.03  &  83.07$\pm$1.69  &  1126.3$\pm$1.4  & 291.7$\pm$0.13  &  64.3  &       \\
WT 1827              &   10 43 02.81 & $-$09 12 40.8 &       V  &     4c &  65 &  4.14 &   8  &     80.47$\pm$2.42  & 0.52$\pm$0.10  &  80.99$\pm$2.42  &  1958.5$\pm$1.4  & 280.0$\pm$0.07  & 114.6  &       \\
DENIS J1048-3956     &   10 48 14.57 & $-$39 56 07.0 &       I  &     4c &  92 &  3.18 &   9  &    246.64$\pm$1.54  & 1.07$\pm$0.18  & 247.71$\pm$1.55  &  1530.4$\pm$1.3  & 229.2$\pm$0.10  &  29.3  &       \\
LHS 300AB            &   11 11 13.68 & $-$41 05 32.7 &       R  &     4c &  65 &  3.18 &  12  &     30.65$\pm$1.84  & 1.65$\pm$0.17  &  32.30$\pm$1.85  &  1249.2$\pm$1.4  & 264.1$\pm$0.10  & 183.3  &     ! \\
LHS 306              &   11 31 08.38 & $-$14 57 21.3 &       R  &     4c &  65 &  3.18 &   8  &     88.20$\pm$1.68  & 1.04$\pm$0.15  &  89.24$\pm$1.69  &  1431.6$\pm$1.6  & 163.2$\pm$0.11  &  76.0  &       \\
LHS 346              &   13 09 20.42 & $-$40 09 27.0 &       V  &     4c &  71 &  3.18 &   9  &     59.78$\pm$1.11  & 1.97$\pm$0.13  &  61.75$\pm$1.12  &  1233.6$\pm$0.9  & 143.6$\pm$0.08  &  94.7  &       \\
ER 2                 &   13 13 09.33 & $-$41 30 39.7 &       V  &     4c &  49 &  2.94 &   8  &     82.28$\pm$1.58  & 1.30$\pm$0.10  &  83.58$\pm$1.58  &  1027.6$\pm$1.8  & 271.7$\pm$0.15  &  58.3  &       \\
LHS 382              &   14 50 41.22 & $-$16 56 30.8 &       I  &     4c &  51 &  3.12 &   8  &     19.92$\pm$2.25  & 0.77$\pm$0.09  &  20.69$\pm$2.25  &  1428.7$\pm$1.8  & 244.0$\pm$0.14  & 327.3  &     ! \\
LHS 406              &   15 43 18.33 & $-$20 15 32.9 &       R  &     5c &  61 &  3.60 &  10  &     45.53$\pm$1.59  & 1.75$\pm$0.28  &  47.28$\pm$1.61  &  1161.4$\pm$1.3  & 194.8$\pm$0.10  & 116.4  &       \\
LHS 423              &   16 35 40.40 & $-$30 51 20.2 &       V  &     4c &  53 &  3.01 &  12  &     49.35$\pm$2.93  & 2.04$\pm$0.23  &  51.39$\pm$2.94  &  1158.2$\pm$2.1  & 223.6$\pm$0.20  & 106.8  &       \\
LHS 440              &   17 18 25.58 & $-$43 26 37.6 &       R  &     4s &  85 &  2.94 &   9  &     35.19$\pm$2.14  & 1.71$\pm$0.47  &  36.90$\pm$2.19  &  1080.4$\pm$2.2  & 233.8$\pm$0.23  & 138.8  &       \\
LTT 6933             &   17 28 07.33 & $-$62 27 14.2 &       R  &     4s &  67 &  2.72 &  10  &     59.37$\pm$1.50  & 1.74$\pm$0.16  &  61.11$\pm$1.51  &   959.3$\pm$1.6  & 197.4$\pm$0.18  &  74.4  &     ! \\
GJ 1226A             &   18 20 57.18 & $-$01 02 58.0 &       I  &     4c &  59 &  3.11 &  10  &     37.31$\pm$5.27  & 1.47$\pm$0.17  &  38.78$\pm$5.27  &  1091.6$\pm$4.4  & 207.8$\pm$0.45  & 133.4  &     ! \\
GJ 1226B             &   18 20 57.18 & $-$01 02 58.0 &       I  &     4c &  59 &  3.11 &  10  &     26.27$\pm$6.14  & 1.47$\pm$0.17  &  27.74$\pm$6.14  &  1095.9$\pm$5.2  & 208.3$\pm$0.52  & 187.3  &     ! \\
LHS 475              &   19 20 54.26 & $-$82 33 16.1 &       V  &     4c+&  66 &  3.75 &   8  &     76.46$\pm$2.00  & 1.88$\pm$0.33  &  78.34$\pm$2.03  &  1267.0$\pm$1.7  & 164.6$\pm$0.13  &  76.7  &       \\
GJ 1252              &   20 27 42.07 & $-$56 27 25.2 &       R  &     4c &  87 &  2.87 &  12  &     48.13$\pm$2.13  & 2.40$\pm$0.18  &  50.53$\pm$2.14  &  1298.7$\pm$2.8  & 161.3$\pm$0.22  & 121.8  &       \\
GJ 1251              &   20 28 03.75 & $-$76 40 15.9 &       R  &     4c &  68 &  2.95 &   7  &     76.96$\pm$2.24  & 2.06$\pm$0.17  &  79.02$\pm$2.25  &  1426.6$\pm$2.2  & 149.6$\pm$0.17  &  85.6  &       \\
LHS 510              &   21 30 47.67 & $-$40 42 29.5 &       R  &     4c &  56 &  3.11 &   8  &     82.44$\pm$2.52  & 1.16$\pm$0.14  &  83.60$\pm$2.52  &  1723.9$\pm$2.4  & 143.1$\pm$0.16  &  97.7  &       \\
LHS 512              &   21 38 43.65 & $-$33 39 55.3 &       V  &     4c &  63 &  3.13 &   8  &     80.34$\pm$2.10  & 1.68$\pm$0.06  &  82.02$\pm$2.10  &  1151.1$\pm$1.9  & 116.7$\pm$0.18  &  66.5  &       \\
LHS 521              &   22 27 59.21 & $-$30 09 32.8 &       R  &     4c &  68 &  3.15 &  11  &     20.60$\pm$1.59  & 1.00$\pm$0.07  &  21.60$\pm$1.59  &  1008.8$\pm$2.2  & 136.8$\pm$0.24  & 221.4  &       \\
GJ 1281              &   23 10 42.16 & $-$19 13 34.9 &       V  &     4c &  62 &  3.12 &   8  &     40.15$\pm$2.31  & 0.99$\pm$0.06  &  41.14$\pm$2.31  &  1428.1$\pm$2.5  & 178.2$\pm$0.15  & 164.5  &       \\
LHS 539              &   23 15 51.61 & $-$37 33 30.6 &       R  &     4c &  54 &  2.89 &   8  &     51.93$\pm$2.01  & 0.92$\pm$0.07  &  52.85$\pm$2.01  &  1311.2$\pm$3.0  &  78.4$\pm$0.22  & 117.6  &       \\
LHS 547              &   23 36 52.31 & $-$36 28 51.8 &       V  &     4c+&  53 &  2.88 &   7  &     85.17$\pm$2.03  & 1.06$\pm$0.05  &  86.23$\pm$2.03  &  1168.9$\pm$2.2  &  87.0$\pm$0.16  &  64.3  &       \\ 
\hline
\multicolumn{15}{c}{Revised Parallaxes}\\				   
\hline
GJ 545               &   14 20 07.36 & $-$09 37 13.4 &       V  &     4c &  49 &  3.16 &   9  &     71.02$\pm$1.38  & 0.50$\pm$0.03  &  71.52$\pm$1.38  &  1020.7$\pm$1.2  & 216.8$\pm$0.14  &  67.6  &     ! \\
GJ 754               &   19 20 47.98 & $-$45 33 29.7 &       V  &     5c & 121 &  4.06 &   9  &    166.76$\pm$1.52  & 2.27$\pm$0.32  & 169.03$\pm$1.55  &  2960.7$\pm$1.1  & 167.5$\pm$0.03  &  83.0  &     ! \\
LHS 500              &   20 55 37.12 & $-$14 03 54.8 &       V  &     4c &  70 &  3.00 &   9  &     81.07$\pm$1.54  & 0.88$\pm$0.05  &  81.95$\pm$1.54  &  1490.4$\pm$1.4  & 108.1$\pm$0.10  &  86.2  &     ! \\
LHS 501              &   20 55 37.76 & $-$14 02 08.1 &       V  &     4c &  70 &  3.00 &   9  &     76.71$\pm$1.49  & 0.88$\pm$0.05  &  77.59$\pm$1.49  &  1492.3$\pm$1.4  & 108.0$\pm$0.09  &  91.2  &     ! \\
\hline
\multicolumn{15}{c}{Parallax Calibration Stars}\\
\hline
LHS 1731             &   05 03 20.08 & $-$17 22 25.0 &       V  &     5c &  84 &  3.72 &   9  &    107.53$\pm$2.15  & 1.17$\pm$0.11  & 108.70$\pm$2.15  &   495.9$\pm$1.5  & 207.8$\pm$0.34  &  21.6  &       \\
LHS 1777             &   05 42 12.70 & $-$05 27 55.6 &       I  &     4c &  45 &  2.70 &   7  &     76.22$\pm$1.77  & 3.50$\pm$0.65  &  79.72$\pm$1.89  &   959.0$\pm$2.2  & 351.2$\pm$0.21  &  57.0  &       \\
GJ 440               &   11 45 42.93 & $-$64 50 29.7 &       V  &     4c &  93 &  3.23 &  10  &    213.41$\pm$2.03  & 1.66$\pm$0.19  & 215.07$\pm$2.04  &  2698.5$\pm$2.0  &  97.6$\pm$0.07  &  59.5  &       \\
GJ 465               &   12 24 52.49 & $-$18 14 32.2 &       V  &     4c &  61 &  3.16 &   6  &    108.45$\pm$2.44  & 1.82$\pm$0.29  & 110.27$\pm$2.46  &  2552.6$\pm$1.9  & 154.4$\pm$0.08  & 109.7  &       \\
Proxima Cen          &   14 29 43.02 & $-$62 40 46.7 &       V  &     5s &  86 &  3.38 &   8  &    772.33$\pm$2.05  & 1.92$\pm$0.38  & 774.25$\pm$2.08  &  3856.0$\pm$2.3  & 281.6$\pm$0.05  &  23.6  &     ! \\
GJ 555               &   14 34 16.82 & $-$12 31 10.2 &       V  &     4s &  69 &  3.39 &   8  &    157.72$\pm$2.62  & 0.74$\pm$0.11  & 158.46$\pm$2.62  &   691.3$\pm$2.2  & 330.3$\pm$0.35  &  20.7  &       \\
GJ 581               &   15 19 26.83 & $-$07 43 20.3 &       V  &     4c & 122 &  2.95 &   8  &    153.50$\pm$2.62  & 1.16$\pm$0.11  & 154.66$\pm$2.62  &  1222.4$\pm$2.8  & 265.7$\pm$0.20  &  37.5  &       \\
\enddata
\tablecomments{Stars with exclamation mark are discussed in the section~\ref{sec:notes}.}
\label{tbl:pi.result}
\end{deluxetable}


\voffset100pt{
\begin{deluxetable}{llrrrccr@{.}lr@{.}lr@{.}lcc}
\rotate
\tablewidth{0pt}
\tabletypesize{\small}
\tablehead{
           \colhead{Name1}   &
           \colhead{Name2}   &
           \colhead{$V$}       &
           \colhead{$R$}       &
           \colhead{$I$}       &
           \colhead{\#}      &
           \colhead{Refs}    &
           \multicolumn{2}{c}{$J$}       &
           \multicolumn{2}{c}{$H$}       &
           \multicolumn{2}{c}{$K_{s}$} &
	   \colhead{Spect.}  &
	   \colhead{Refs}    \\
	
	   \colhead{(1)}     &
	   \colhead{(2)}     &
	   \colhead{(3)}     &
	   \colhead{(4)}     &
	   \colhead{(5)}     &
	   \colhead{(6)}     &
	   \colhead{(7)}     &
           \multicolumn{2}{c}{(8)}     &
           \multicolumn{2}{c}{(9)}     &
           \multicolumn{2}{c}{(10)}     &
	   \colhead{(11)}     &
	   \colhead{(12)}     
           }
\startdata
  GJ 1025         & LHS 130       &  13.35 &    12.08 &    10.52 &    2 &    &   9&04   &    8&49   &    8&22   & M3.5 V  &6    \\ 
  GJ 1050         & LHS 157       &  11.79 &    10.68 &     9.35 &    1 &    &   8&06   &    7&54   &    7&31   & M3.5 V  &6    \\
  LHS 158         & L 298-71      &  13.64 &    12.66 &    11.60 &    2 &    &  10&43   &    9&94   &    9&73   & ....    &     \\
  GJ 1068         & LHS 22        &  13.62 &    12.21 &    10.44 &    2 &    &   8&75   &    8&21   &    7&90   & M4.5 V  &6    \\
  LHS 193 A       &               &  11.66 &    10.85 &    10.09 &    3 &    &   9&18   &    8&55   &    8&43   & ....    &     \\
  LHS 193 B       &               &  17.66 &    17.03 &    16.08 &    3 &    &  15&98C  &   15&79C  &   15&29C  & ....    &     \\
  LHS 225 A       &               &  12.87 &    11.85 &    10.51 &    1 &    &   8&61JD &    8&07JD &    7&87JD & ....    &     \\
  LHS 225 B       &               &  13.02 &    11.99 &    10.63 &    1 &    &    &...  &     &...  &     &...  & ....    &     \\
  GJ 1118         & LHS 258       &  13.79 &    12.56 &    10.95 &    2 &    &   9&41   &    8&86   &    8&59   & M3.0 V  &6    \\
  GJ 1123         & LHS 263       &  13.16 &    11.86 &    10.16 &    3 &    &   8&33   &    7&77   &    7&45   & M4.5 V  &8    \\
  GJ 1128         & LHS 271       &  12.74 &    11.39 &     9.65 &    3 &    &   7&95   &    7&39   &    7&04   & M4.5 V  &8    \\
  GJ 1129         & LHS 273       &  12.39 &    11.20 &     9.64 &    2 &    &   8&12   &    7&54   &    7&26   & M3.5 V  &8    \\
  WT 248          &               &  14.52 &    13.40 &    11.95 &    2 & 9  &  10&56   &   10&10   &    9&87   & M3.0 V  &8    \\
  LHS 281         & GJ 1132       &  13.49 &    12.26 &    10.69 &    2 &    &   9&25   &    8&67   &    8&32   & M3.5 V  &6    \\
  WT 1827         &               &  15.11 &    13.57 &    11.59 &    2 & 9  &   9&67   &    9&10   &    8&73   & ....    &     \\
\multicolumn{2}{l}{DENI1048-3956} &  17.39 &    15.06 &    12.57 &    3 &    &   9&54   &    8&91   &    8&45   & M8.5 V  &4    \\
  LHS 300 A       & L 395-13      &  13.18 &    12.28 &    11.49 &    1 &    &  10&48J  &   10&01J  &    9&80J  & K4.0 VJ &3    \\
  LHS 300 B       &               &  17.25 &    16.62 &    15.70 &    1 &    &    &...  &     &...  &     &...  & ....    &     \\
  LHS 306         &               &  14.19 &    12.80 &    11.05 &    2 &    &   9&36   &    8&76   &    8&50   & M4.5 V  &6    \\
  LHS 346         &               &  12.86 &    11.73 &    10.24 &    2 &    &   8&79   &    8&24   &    7&99   & M3.5 V  &6    \\
  ER2             &               &  12.91 &    11.61 &     9.96 &    1 & 9  &   8&29   &    7&68   &    7&41   & ....    &     \\
  LHS 382         &               &  15.30 &    14.61 &    13.17 &    2 &10  &  11&85   &   11&38   &   11&11   & ....    &     \\
  LHS 406         & GJ 2116       &  13.06 &    12.07 &    10.93 &    2 &    &   9&78   &    9&23   &    9&02   & M1.0 V  &3    \\
  LHS 423         & L 555-14      &  12.66 &    11.58 &    10.16 &    1 &    &   8&89   &    8&36   &    8&08   & M3.0 V  &6    \\
  LHS 440         & L 413-156     &  12.98 &    11.98 &    10.86 &    2 &    &   9&70   &    9&13   &    8&95   & M1.0 V  &3    \\
  LTT 6933        & LHS 3292      &  12.74 &    11.54 &     9.95 &    1 &    &   8&42D  &    7&85D  &    7&57D  & ....    &     \\
  GJ 1226 A       & LHS 463 A     &  13.07 &    11.98 &    10.56 &    1 &    &   8&75J  &    8&19J  &    7&95J  & M3.5 VJ &6    \\
  GJ 1226 B       & LHS 463 B     &  13.16 &    12.07 &    10.63 &    1 &    &    &...  &     &...  &     &...  & ....    &     \\
  LHS 475         & L 22-69       &  12.68 &    11.50 &    10.00 &    2 &    &   8&56   &    8&00   &    7&69   & M3.0 V  &6    \\
  GJ 1252         & LHS 492       &  12.20 &    11.19 &     9.93 &    1 &    &   8&70   &    8&16   &    7&92   & M2.5 V  &6    \\
  GJ 1251         & LHS 493       &  13.96 &    12.76 &    11.11 &    1 &    &   9&36   &    8&88   &    8&60   & M4.5 V  &6    \\
  LHS 510         & L 425-35      &  13.12 &    11.92 &    10.34 &    1 &    &   8&87   &    8&42   &    8&13   & M1.5 V  &6    \\
  LHS 512         & L 570-29      &  12.55 &    11.38 &     9.89 &    1 &    &   8&44   &    7&84   &    7&57   & M3.5 V  &6    \\
  LHS 521         &               &  14.68 &    13.84 &    13.10 &    1 &    &  12&13   &   11&66   &   11&46   & ....    &     \\
  GJ 1281         & LHS 538       &  12.45 &    11.42 &    10.19 &    1 &    &   8&98   &    8&46   &    8&23   & M2.5 V  &6    \\
  LHS 539         &               &  14.98 &    13.66 &    11.96 &    2 &    &  10&40   &    9&87   &    9&59   & ....    &     \\
  LHS 547         & L 504-27      &  13.76 &    12.46 &    10.79 &    3 &    &   9&19   &    8&67   &    8&42   & M4.5 V  &6    \\
\hline                                                                                                                                
  GJ 545          & LHS 369       &  12.84 &    11.69 &    10.15 &    2 &    &   8&74   &    8&19   &    7&98   & M3.5 V  &6    \\
  GJ 754          & LHS 60        &  12.26 &    10.93 &     9.24 &    1 & 2  &   7&66   &    7&13   &    6&85   & M4.5 V  &6    \\
  LHS 500         & GJ 810 B      &  14.61 &    13.19 &    11.38 &    1 & 2  &   9&72   &    9&22   &    8&92   & M5.0 V  &6    \\
  LHS 501         & GJ 810 A      &  12.48 &    11.22 &     9.61 &    1 & 2  &   8&12   &    7&64   &    7&37   & M4.0 V  &6    \\
\hline                                                                                                                              
  LHS  1731       &               &  11.69 &    10.59 &     9.16 &    1 &10  &   7&82   &    7&24   &    6&94   & M3.0 V  &8    \\
  LHS  1777       &               &  15.29 &    13.81 &    11.92 &    1 & 5  &  10&21   &    9&69   &    9&37   & M5.5 V  &6    \\
  GJ 440          & LHS 43        &  11.50 &    11.33 &    11.19 &    3 & 1  &  11&19   &   11&13   &   11&10   & DQ6.0   &11   \\
  GJ 465          & LHS 45        &  11.27 &    10.23 &     8.92 &    1 & 2  &   7&73   &    7&25   &    6&95   & M2.0 V  &6    \\
  Proxima Cen     & LHS 49        &\nodata & \nodata  &  \nodata &\nodata& 2 &   5&36   &    4&84   &    4&38   & M5.5 V  &6    \\
  GJ 555          & LHS 2945      &  11.30 &    10.05 &     8.43 &    1 &10  &   6&84   &    6&26   &    5&94   & M3.5 V  &8    \\
  GJ 754          & LHS 394       &  10.55 &     9.44 &     8.04 &    2 & 2  &   6&71   &    6&10   &    5&84   & M3.0 V  &6    \\ 
\enddata

\tablecomments{C: affected by confusion with another nearby source; D:
affected by a nearby diffraction spike; J: joined (combined)
photometry}
\tablerefs{(1): \citealt{Bergeron2001}; (2): \citealt{Bessel1990}; (3):
\citealt{Bid1985}; (4): \citealt{Henry2004}; (5): \citealt{Har1993}; (6):
\citealt{Haw1996}; (7): \citealt{Henry1994}; (8): \citealt{Henry2002}; (9):
\citealt{Pat1998}; (10): \citealt{Wei1996}; (11): \citealt{McCook1999}.}
\label{tbl:phot.result}
\end{deluxetable}
}


\begin{thebibliography}{}

\bibitem[Bakos, Sahu, \& N{\'e}meth (2002)]{Bakos2002} Bakos, G.~{\'
A}., Sahu, K.~C., \& N{\'e}meth, P.\ 2002, \apjs, 141, 187

\bibitem[Benedict et al.(1999)]{Benedict1999} Benedict, G.~F., et al.\
1999, \aj, 118, 1086

\bibitem[Bergeron et al.(1995)]{Bergeron1995} Bergeron, P., 
Wesemael, F., \& Beauchamp, A.\ 1995, \pasp, 107, 1047 

\bibitem[Bergeron (2001)]{Bergeron2001b} Bergeron, P., 2001, \apj, 558,
369

\bibitem[Bergeron, Leggett, \& Ruiz (2001)]{Bergeron2001} Bergeron,
P., Leggett, S.~K., \& Ruiz, M.~T.\ 2001, \apjs, 133, 413

\bibitem[Bertin \& Arnouts(1996)]{sextractor} Bertin, E.~\& Arnouts,
S.\ 1996, \aaps, 117, 393

\bibitem[Bessel(1990)]{Bessel1990} Bessel, M.~S.\ 1990,
\aaps, 83, 357

\bibitem[Bidelman(1985)]{Bid1985} Bidelman, W.~P.\ 1985, \apjs, 59,
197


\bibitem[Costa et al.(2005)]{Costa2004} Costa, E. et al, 2005, AJ, in preparation

\bibitem[Dahn et al.(2002)]{Dahn2002} Dahn, C.~C., et al.\ 2002, \aj,
124, 1170

\bibitem[Deacon \& Hambly(2001)]{Deacon2001} Deacon, N.~R.~\& Hambly,
N.~C.\ 2001, \aap, 380, 148

\bibitem[ESA(1997)]{Hipparcos} The {\it Hipparcos} and Tycho
Catalogues, \ 1997, ESA SP-1200 (Noordwijk: ESA)

\bibitem[Eichhorn(1974)]{Eichhorn1974} Eichhorn, H., 1974, Astronomy
\& of star positions; a critical investigation of star catalogues, the
\& methods of their construction, and their purpose, Ungar, New York,
1974

\bibitem[Gizis(1997)]{Gizis1997} Gizis, J.~E.\ 1997, \aj, 113, 806

\bibitem[Gizis \& Reid(1997)]{Gizis1997b} Gizis, J.~E. \& Reid, N.I.,
\ 1997, \pasp, 109, 849

\bibitem[Gizis(1998)]{Gizis1998} Gizis, J.~E.\ 1998, \aj, 115, 2053


\bibitem[McCook \& Sion(1999)]{McCook1999} McCook, G.~P., \& Sion, 
E.~M.\ 1999, \apjs, 121, 1 

\bibitem[Graham(1982)]{Graham1982} Graham, J.~A.\ 1982, \pasp, 94, 244

\bibitem[Harrington et al.(1993)]{Har1993} Harrington, R.~S., et al.\
1993, \aj, 105, 1571

\bibitem[Hambly et al.(1999)]{Hambly1999} Hambly, N.~C., et al.\ 1999,
\mnras, 309, L33

\bibitem[Hambly et al.(2004)]{Hambly2004} Hambly, N.~C., Henry, 
T.~J., Subasavage, J.~P., Brown, M.~A., \& Jao, W.\ 2004, \aj, 128, 437 

\bibitem[Hansen(1998)]{Hansen1998} Hansen, Brad~M.~S.\ 1998, \nat,
394, 860

\bibitem[Hawley, Gizis, \& Reid(1996)]{Haw1996} Hawley, S.~L., Gizis,
J.~E., \& Reid, I.~N.\ 1996, \aj, 112, 2799

\bibitem[Henry et al.(2003)]{Henry2003} Henry, T.~J., Backman, D.~B.,
Blackwell, J., Okimura, T., \& Jue, S.\ 2003, in {\it The Future of
Small Telescopes In The New Millennium. Volume III --- Science in the
Shadows of Giants}, ed. T.D. Oswalt, Astrophysics and Space Science
Library, Volume 289, Kluwer Academic Publishers, Dordrecht, 2003.,
p.111

\bibitem[Henry, Kirkpatrick, \& Simons(1994)]{Henry1994} Henry, T.~J.,
Kirkpatrick, J.~D., \& Simons, D.~A.\ 1994, \aj, 108, 1437

\bibitem[Henry et al.(1997)]{Henry1997} Henry, T.~J., Ianna, P.~A.,
Kirkpatrick, J.~D., \& Jahrei{\ss}, H.\ 1997, \aj, 114, 388

\bibitem[Henry et al.(2002)]{Henry2002} Henry, T.~J., Walkowicz,
L.~M., Barto, T.~C., \& Golimowski, D.~A.\ 2002, \aj, 123, 2002

\bibitem[Henry et al.(2004)]{Henry2004} Henry, T.~J., Subasavage,
J.~P., Brown, M.~A., Beaulieu, T.~D., Jao, W., \& Hambly, N.~C.\ 2004,
\aj, 128, 2460

\bibitem[Ianna et al.(1994)]{Ianna1994} Ianna, P.~A., Begam, M.~C., \&
Mullis, C.~R.\ 1994, Bulletin of the American Astronomical Society,
26, 1347

\bibitem[Ianna et al.(1996)]{Ianna1996} Ianna, 
P.~A., Patterson, R.~J., \& Swain, M.~A.\ 1996, \aj, 111, 492 

\bibitem[Jao et al.(2003)]{Jao03} Jao, W.-C, Henry, T.~J., Subasavage,
J.~P., Bean, J.~L., Costa, E., Ianna, P.~A., \& M{\' e}ndez, R.~A.\
2003, \aj, 125, 332

\bibitem[Jefferys, Fitzpatrick \& McArthur(1987)]{Jefferys1987}
Jefferys, W.~H., Fitzpatrick, M.~J., \& McArthur, B.~E.\ 1987,
Celestial Mechanics, 41, 39

\bibitem[Landolt(1992)]{Landolt1992} Landolt, A.~U.\ 1992, \aj, 104,
340

\bibitem[Luyten(1979)]{LHS} Luyten, W.~J.\ 1979, LHS Catalog,
Minneapolis, University of Minnesota, 1979, 2nd ed.

\bibitem[Luyten(1980)]{NLTT} Luyten, W.~J.\ 1980, NLTT Catalog,
Minneapolis, University of Minnesota, 1980

\bibitem[Monet et al.(1992)]{Monet1992} Monet, D.~G., Dahn, C.~C.,
Vrba, F.~J., Harris, H.~C., Pier, J.~R., Luginbuhl, C.~B., \& Ables,
H.~D.\ 1992, \aj, 103, 638

\bibitem[Patterson, Ianna, \& Begam (1998)]{Pat1998} Patterson, R.~J.,
Ianna, P.~A., \& Begam, M.~C.\ 1998, \aj, 115, 1648

\bibitem[Tinney(1993)]{Tinney1993} Tinney, C.~G.\ 1993, \aj, 105, 
1169 

\bibitem[Tinney et al.(1995)]{Tinney1995} Tinney, 
C.~G., Reid, I.~N., Gizis, J., \& Mould, J.~R.\ 1995, \aj, 110, 3014 

\bibitem[Tinney et al.(2003)]{Tinney2003} Tinney,
C.~G., Burgasser, A.~J., \& Kirkpatrick, J.~D.\ 2003, \aj, 126, 975

\bibitem[Stone(1996)]{Stone1996} Stone, R.~C.\ 1996, \pasp, 108, 1051

\bibitem[Stone(2002)]{Stone2002} Stone, R.~C.\ 2002, \pasp, 114, 
1070 

\bibitem[Subasavage et al.(2005)]{Subasavage2005} Subasavage, J.~P., AJ, 2005, in press

\bibitem[van Altena(1974)]{Altena1974} van Altena, W.~F.\ 1974, \aj,
79, 826

\bibitem[van Altena et al.(1986)]{Altena1986} van Altena, W.~F., Auer,
L.~H., Mora, C.~L., \& Vilkki, E.~U.\ 1986, \aj, 91, 1451

\bibitem[van Altena et al.(1988)]{Altena1988} van Altena, W.~R. et
al., "Parallax calibration of the population II main sequence, II. The
effect of changes in the corrections to absolute parallax", ,
Calibration of stellar ages, 1988, Philip, A. G. D., 7, 175--184,
Schenectady, N.Y.

\bibitem[van Altena et al.(1995)]{YPC} van Altena, W.~F., Lee, J.~T.,
\& Hoffleit, D.\ 1995, The General Catalogue of Trigonometric Stellar
Parallaxes (4th ed.; New Haven: Yale Univ. Obs.)

\bibitem[Vilkki(1984)]{Vilkki1984} Vilkki, E.~U.\ 1984, \pasp, 96, 
161 

\bibitem[Vrba et al.(2004)]{Vrba2004} Vrba, F.~J., et al.\ 2004, 
\aj, 127, 2948 

\bibitem[Weis(1996)]{Wei1996} Weis, E.~W.\ 1996, \aj, 112,
2300

\end{thebibliography}
\end{document}